\documentclass{jfm}

\usepackage{graphicx}
\usepackage{newtxtext}
\usepackage{newtxmath}
\usepackage{natbib}
\usepackage{hyperref}
\usepackage{color}
     \usepackage{xcolor}

\hypersetup{
    colorlinks = true,
    urlcolor   = blue,
    citecolor  = black,
}

\newcommand{\RomanNumeralCaps}[1]
\linenumbers
\definecolor{g-blue}{rgb}{0.83,0.95,1}
\definecolor{g-yellow}{rgb}{1,1,0.7}
\definecolor{g-green}{rgb}{0.9,1,0.9}
\definecolor{green}{rgb}{0,0.6,0}
\definecolor{cyan}{rgb}{0,0.7,0.7} 
\definecolor{grey}{rgb}{0.4 ,0.4 ,0.4 }
\definecolor{brown}{rgb}{0.6 ,0  ,0.8 }

\def\g-blue#1{\textcolor{g-blue}{#1}}

\usepackage{bm}

\DeclareMathOperator{\sign}{sign}

\def\bp{\mathbf{p}}
\def\bk{\mathbf{k}}

\def\mR{\mathcal{R}}
\def\mI{\mathcal{I}}
\def\mL{\mathcal{L}}

\title{Downscale energy fluxes in scale invariant \\ oceanic internal wave turbulence}

\author{Giovanni Dematteis\aff{1}
 \and Yuri V. Lvov\aff{1}}

\affiliation{\aff{1} Department  of Mathematical  Sciences, Rensselaer
  Polytechnic Institute,\\$\;\,$110 8th St, Troy, NY 12180, US}

\begin{document}
\maketitle

\begin{abstract}

We analyze analytically and numerically the scale invariant stationary
solution to the internal wave kinetic equation. Our analysis of the
resonant energy transfers shows that the leading order contributions
are given (i) by triads with extreme scale separation and (ii) by
triads of waves that are quasi-colinear in the horizontal plane. The
contributions from other types of triads is found to be subleading.
We use the modified scale invariant limit of the Garrett and Munk
spectrum of internal waves to calculate the magnitude of the energy
flux towards high wave numbers in both the vertical and the horizontal
directions. Our results compare favorably with the finescale
parametrization of ocean mixing that was proposed
in~[\cite{polzin1995finescale}].
\end{abstract}


\section{Introduction}\label{sec:0}
        {\it Internal waves} are the gravity waves that oscillate in
        the bulk of the stratified ocean due to the modulation of
        surfaces of constant density.  Internal waves are ubiquitous
        in the ocean, contain a large amount of energy, and affect
        significantly the processes involved in water mixing and
        transport.
Understanding the role played by internal gravity waves in the energy
budget of the oceans represents a major challenge in physical
oceanography, intimately related to the quantification of ocean
mixing~[\cite{Ferrari,polzin2014finescale}]. Internal waves constitute
a highly complex problem, involving scales from few meters to hundreds
of kilometers, and periods ranging from a few minutes up to days. The
processes that supply energy to and remove energy from internal waves
include interactions with surface gravity waves, mesoscale eddies,
scattering of the tidal flow from the bottom topography, overturning
of wave fronts and wave breaking. These pumping and damping processes
are characterized by vastly different spatial and temporal
scales. Despite this enormous complexity, the spectral energy density
of internal waves is thought to be pretty universal, and is given by
what is now called Garrett and Munk spectrum of internal waves. First
proposed in 1972~[\cite{GM72}], then subject to subsequent
revisions,~[\cite{GM75,GM76,GM79}], the Garrett and Munk spectrum (GM,
from now on referring to the 1976 version) has since become the
accepted default choice to quantify the oceanic internal wave field.
Since then substantial deviations, both seasonal and regional, were
documented (notably near
boundaries,~[\cite{WW79,polzin2004heuristic,regional}], and at the
equator,~[\cite{E85}]). Nonetheless, the GM spectrum has survived to
our day as the standard for inter-comparison of different data sets,
to the amazement of Chris Garrett and Walter Munk
themselves~[\cite{von2010internal}], providing the baseline for
generalizations that try to account for the observed
variability~[\cite{regional}].

The spectral energy fluxes in the oceanic internal wave field have
been a subject of intense investigation in the last four
decades~[\cite{Muller86,McComas1977,Olbers1973,polzin1995finescale}]. Understanding
these energy fluxes is crucial for climate modeling and predictions,
since internal waves are not resolved in Global Circulation Models
(GCM) and they are replaced by simple phenomenological
formulas~[\cite{mackinnon2017climate}] . One of such broadly used
expressions is the finescale parametrization formula derived
in~[\cite{polzin1995finescale}].

In this paper we use the wave turbulence theory for internal gravity
waves developed in~[\cite{LT,LT2,iwthLPTN,LY}] and reviewed
in~[\cite{regional}] to analyze these energy fluxes towards high wave
numbers. We assume that the spectral energy density of internal waves
is given by a simple scale invariant solution, that was found
in~[\cite{iwthLPTN}], (hereafter, {\it the {\it convergent stationary}
  solution} of the internal wave kinetic equation, Eq. (\ref{eq:3.69})
in the body of the paper).  Interestingly, this scale invariant
spectrum is close to the scale invariant limit of the famous Garrett
and Munk (GM) spectrum of internal waves~[\cite{GM72,GM76,GM79}], see
Eq.~\eqref{eq:4} in the body of the paper. Thus, slightly adjusting
the GM spectrum so that its scale invariant limit matches the
power-law behavior of the convergent stationary solution, we compute
the energy flux via the collision integral of the wave kinetic
equation. This collision integral contains complete information
concerning resonant spectral energy transfers. The computation of the
flux is performed numerically.  Our expression for these energy fluxes
compares favorably with the finescale parametrization formula put
forward in~[\cite{polzin1995finescale}].

To characterize the energy fluxes towards high wave numbers we
investigate the formation of the stationary wave spectra in the
kinetic equation. We therefore analyze and classify the contributions
of the various resonant triads that contribute to the kinetic
equation.  The importance of triads with extreme scale separation were
previously identified in the literature~[\cite{McComas1977}] and named
Induced Diffusion (ID), Parametric Subharmonic Instability (PSI) and
Elastic Scattering (ES).  In addition, we point out an additional
class of important interactions that are colinear, or almost so, in
the horizontal plane, which appear to contribute
  significantly to the formation of the stationary state and the
fluxes of energy.  

The paper is written as follows.  In Section \ref{sec:Background} we
give the reader the relevant background along with necessarily brief
literature review.  In Section \ref{sec:1} we analyze the convergence
conditions of the collision integral at the infrared and the
ultraviolet limits. We perform a rigorous numerical integration
paying special attention to accurately integrate integrable
singularities of the kinetic equation kernel. We also analyze in
details the nature of the interacting triads contributing to the
stationary scale invariant solution of the kinetic
equation. An analytical and numerical analysis of these various
contributions is presented in Section \ref{sec:2}, highlighting the
main physical mechanisms at play.  In Section \ref{sec:3} we compute
the energy fluxes toward the small scales and thereby quantify the
total dissipated energy. Finally, we summarize our results in Section
\ref{sec:conclusions}.

\section{Background material}\label{sec:Background}
\subsection{Internal waves and the Garrett and Munk spectrum}

Garrett and Munk have observed that the internal wave spectrum is
{\it{separable}} in frequency-vertical wave numbers. In other words it
can be accurately represented as a {\it{product}} of a function of
frequency and a function of vertical wave number.  The GM
energy spectrum is therefore represented in the two-dimensional domain
of vertical wavenumber $m$ with $m\in [m_{\min},m_{\max}]$ and
frequency $\sigma$ with $\sigma\in[f,N]$ as
\begin{eqnarray}\label{eq:1a}
	&\qquad e_{\rm GM}(m,\sigma) = N_0 N b^2 E A(m) B(\sigma)\,,\\
\label{eq:1b}
	&A(m)=\frac{2}{\pi}\frac{m_\star}{m_\star^2+m^2}\,,\qquad
        B(\sigma) = \frac{2f}{\pi}\frac{1}{\sigma\sqrt{\sigma^2-f^2}}
        \,, \qquad m_\star = \frac{3\pi}{b}{\color{black}\frac{N}{N_0}}\,,
\end{eqnarray}
normalized in such a way that $\int_{m_{\min}}^{m_{\max}}A(m) dm\simeq
\int_0^\infty A(m)dm =1$, $\int_{f}^{N}B(\sigma) d\sigma \simeq
\int_0^\infty B(\sigma)d\sigma =1$, and the total energy density (per
unit mass) is therefore given by $N_0N b^2 E$, in units of ${\rm
  J}\,{\rm kg}^{-1}$. Here $N$ and $N_0=0.00524\;{\rm s}^{-1}$ are
respectively the buoyancy frequency and the reference buoyancy
frequency, $f=2\times7.3\times10^{-5} \sin(l) \;{\rm s}^{-1}$ is the
Coriolis parameter computed at latitude $l=32.5^\circ$, $b=1300\;{\rm
  m}$ is the scale height of the ocean, $E=6.3\times 10^{-5}$ is the
GM specification of the nondimensional energy level, and $m_\star$ is
a reference vertical wave number. Furthermore, $m_{\rm min} =2\pi (2600 \text{
  m})^{-1}\,,$ $ m_{\rm max} =2\pi (10 \text{ m})^{-1}$ are the
physical cutoffs imposed by the ocean depth and by wave breaking,
respectively.

\subsection{Wave-turbulence interpretation of the GM spectrum}
\noindent Despite the GM far reaching combination of simplicity and
descriptive power of available field measurements, its
phenomenological nature does not necessarily provide an explanation to
the underlying physics. Since the 1970s, the concept of nonlinear
interactions has become the {\it leit motiv} in the search for a
physical interpretation of the GM spectrum starting from the primitive
equations of a stratified ocean,
[\cite{McComas1977,MM81,M86,Muller86,Olbers1973,Olbers1976,pelinovsky1977weak,caillol2000kinetic,Voronovich,LT}]. The
quadratic nonlinearity in the primitive fluid equations and a
dispersion relation allowing for three-wave interactions imply that
internal waves interact through triads. In a weakly nonlinear regime,
three-wave resonant interactions are responsible for slow, net energy
transfers between different wave numbers, [\cite{davis2020succession}]. This process can be
described by a wave kinetic equation, the evolution equation of the
action spectrum of the internal wavefield,
[\cite{hasselmann1966feynman,ZLF,NazBook}]. In the present paper
we use the three-dimensional wave number domain $\bp=(\bk,m)$, where
$\bk$ and $m$ are the horizontal and the vertical wave numbers,
respectively. Note that $\bk$ is a two-dimensional horizontal wave
vector and we define its norm as $k:=|\bk|$. The dispersion relation
of internal gravity waves is given by
$\sigma^2_{\bf p} = f^2 + N^2\frac{k^2}{m^2}$.
%
which can be used to switch from one domain to the other, since only
two of the three variables $k$, $m$ and $\sigma$ are independent. The
action spectrum is related to the energy spectrum (now both intended
as three-dimensional spectra) via $e(\bk,m)=\sigma n(\bk,m)$, where
for simplicity we use the quantities in brackets to specify the domain
of dependence of the quantity of interest. We assume horizontal {\color{black}and vertical}
isotropy, so that we have $e(m,\sigma)={\color{black}4}\pi k
e(\bk,m)\left(\frac{d\sigma}{dk}\right)^{-1}$, after integrating over
the horizontal azimuthal angle {\color{black}and considering a positive definite $m$}. Considering the scale invariant (or {non-rotating}) limit
$f\ll\sigma\ll N$, which yields the scale-invariant dispersion
relation (here defined as the positive branch) $\sigma = N k/|m|\,$,
Eq.~\ref{eq:1a} transforms into
\begin{equation}\label{eq:4}
	n_{\rm GM}(\bk,m) = \frac{{\color{black}1}}{\pi^3} \frac{E b^2 {\color{black}N_0} f m_\star}{{\color{black}N}} k^{-4}\,,
\end{equation}
which represents the {non-rotating} limit of the GM
three-dimensional action spectrum, in the horizontal wave number -
vertical wave number domain.

{The wave turbulence theory for internal waves was
  revisited with the generalized {\it random phase and amplitude}
  formalism ([\cite{X2,X3,NazBook}]) in the series of works} [\cite{LT,LPT,iwthLPTN}], where the internal wave
kinetic equation was derived starting from the primitive equations of
motion in hydrostatic balance by using isopycnal coordinates.
A detailed review is found in the introductory paragraphs of
[\cite{iwthLPTN}], and will not be repeated here. The main steps can
be schematized as follows: {\bf (i) }The primitive equations of a
vertically stratified ocean in hydrostatic balance and with no
background rotation are rewritten in isopycnal coordinates under the
Boussinesq approximation. The scale invariant limit of the dispersion
relation in the new variables reads (with no background rotation)
\begin{equation}\label{eq:5}
	\sigma(\bk,m) =\frac{g}{\rho_0 N} \frac{k}{|m|}\,,
\end{equation}
where $g$ is the acceleration of gravity, $\rho_0$ is the reference
density and the vertical wave number $m$ is now an inverse density.
{\bf{(ii)}} In the isopycnal formulation the equations of motion are
reduced to Hamiltonian form for the two conjugate fields $\phi$ and
$\Pi$, the velocity potential and the normalized differential layer
thickness. {\bf{(iii)}} The machinery of wave turbulence is applied by
switching to Fourier space and introducing the complex canonical
normal variables $c_\bp$ and $c_{-\bp}^*$, representing complex
amplitudes of the normal modes of the system. Under the assumption of
spatial homogeneity, the action spectral density is defined as
\begin{equation}\label{eq:5b}
	\langle c_{\bp_1} c_{\bp_2}^* \rangle = n_{\bp_1}\delta_{\bp_1-\bp_2}\,,
\end{equation}
where $\delta(\cdot)$ is a Dirac delta, and the angular brackets
denote averaging on a suitably defined statistical ensemble: under the
standard assumptions of random phases and amplitudes
[\cite{ZLF,NazBook}], in the joint limit of large box and small
nonlinearity the following wave kinetic equation is derived, assuming
isotropy in the horizontal plane (for simplicity, here written in the
non-rotating limit:
\begin{equation}\label{eq:6}
	\frac{\partial n_\bp}{\partial t} = \frac{8\pi}{k}\int \left(
        f^\bp_{12} |V^\bp_{12}|^2\delta_{m-m_1-m_2}
        \delta_{\sigma_\bp-\sigma_1-\sigma_2} \frac{k k_1
          k_2}{\Delta_{\bp12}} - (0\leftrightarrow1) -
        (0\leftrightarrow2)\right) dk_1 dk_2 dm_1dm_2\,,
\end{equation}
%
where $n_\bp=n(\bk,m;t)$ is the three-dimensional action spectrum
defined in~\eqref{eq:5b}, $f^\bp_{12}=n_1n_2-n_\bp(n_1+n_2)$,
$V^\bp_{12}$ is the matrix element describing the magnitude of
nonlinear interactions between the triad of wave numbers $\bp$,
$\bp_1$ and $\bp_2$, given below by (\ref{eq:12}). Furthermore the two
delta functions impose the conservation of vertical momentum and
energy in each three-wave interaction. The $\Delta_{\bp12}$, given
by (\ref{Delta}), is a factor coming from integration of the horizontal
momentum delta function, proportional to the area of the triangle with
sides $k$, $k_1$ and $k_2$.  {\bf(iv)} The wave kinetic
equation~\eqref{eq:6} with dispersion relation~\eqref{eq:5} is fully
scale invariant and the consequent theory of power-law spectra,
[\cite{ZLF}], was worked out in [\cite{iwthLPTN}]. Assuming a
solution of type
\begin{equation}\label{eq:9}
	n_{\bp}\propto k^{-a} |m|^{-b}\,,
\end{equation}
the stationary solution corresponding to constant energy flux,
i.e. the Kolmogorov-Zakharov (KZ) spectrum, can be derived by
Zakharov-Kuznetsov conformal mapping~[\cite{ZLF}] yielding $a=7/2$, $b=1/2$:
\begin{equation}\label{PelinovskyRaevsky}
  n^{PR}(k,m) = k^{-\frac{7}{2}}m^{-\frac{1}{2}}\,.
\end{equation}
Such a solution was derived in [\cite{pelinovsky1977weak}] and again
in [\cite{LT}] and is known as the Pelinovski-Raevski (PR)
spectrum. {\bf{(v)}} A KZ spectrum is a valid solution of the wave
kinetic equation if and only if the locality conditions are satisfied,
i.e. when the collision integral on the r.h.s. of the wave kinetic
equation converges. It turns out that this is not the case for the PR
spectrum; more precisely, in the $a-b$ {power-law} space [\cite{iwthLPTN}] found
that the collision integral {\it converges} only on the segment $b=0,
3.5<a<4$. (vi) On this convergence segment, it was shown by direct
numerical integration that the collision integral is zero for $a\simeq
3.7$, locating the scale invariant stationary solution of the wave
kinetic equation at the point $a=3.7, b=0$. Since this is not far from
the $a=4, b=0$ point of Eq.~\eqref{eq:4}, such a solution has
therefore been put forward as the possible theoretical explanation of
the GM spectrum provided by wave turbulence.

In the present work we use the wave turbulence kinetic equation to
analyze how the stationary scale invariant internal wave spectrum is
formed, and we calculate the corresponding energy fluxes related to
this spectrum.  This quantity is modelled phenomenologically, as
interpretation of the available data, by what is known as the {\it
  finescale parametrization} of the oceanic turbulent mixing,
[\cite{polzin1995finescale,polzin2014finescale,whalen2012spatial,mackinnon2017climate,liang2018assessment}],
and represents a fundamental building block of the Global Circulation
Models.

\section{The convergent stationary solution of the wave kinetic equation}\label{sec:1}
Our starting point is the scale invariant wave kinetic
equation~\eqref{eq:6}. In [\cite{iwthLPTN}] the locality conditions on
the exponents $a,b$ of Eq~\eqref{eq:9} were computed, yielding
convergence conditions $b=0, 3.5<a<4$. We repeated those calculations
confirming that for $b\neq 0$ the collision integral is divergent,
i.e. corresponds to interactions that are {\it nonlocal} in Fourier
space, because of divergence in the infrared or the ultraviolet
limits, or both. In the present paper we focus our attention to the
case $b=0$. We recompute the leading order of the integrand at the
boundaries of the kinematic box, showing that the infrared convergence
condition gives $a<4$ and the ultraviolet convergence condition gives
$a>3$. The combination of the two conditions yields a convergence
segment $3<a<4$, different from the condition $3.5<a<4$ found in
[\cite{iwthLPTN}]. The correction to the previous result is due to a
second exact cancellation in the ultraviolet divergence, previously
undetected. We use a rigorous numerical procedure, with details in
{\it Supplementary materials} [\cite{GiovanniSupplemental}], to
compute the integrable singularities accurately by exploiting the
analytical knowledge of the leading order terms. By direct numerical
computation, we numerically confirm that on the convergence segment
the collision integral tends to $-\infty$ as $a\to3^+$, due to the
ultraviolet divergence, and it tends to $+\infty$ as $a\to4^-$, due to
the infrared divergence. Moreover, it is monotonically increasing with
$a$, crossing zero at $a=3.69$. {Numerical convergence is checked to a
high degree of accuracy}. The independent computation of the convergent
stationary spectrum $a=3.69,b=0$ is the first important result of the
paper, confirming the previous result in [\cite{iwthLPTN}], although a
correction to the convergence segment has been made.
\subsection{Locality conditions}\label{sec:loccond}
\noindent Let us consider the wave kinetic equation of internal
gravity waves in a non-rotating frame, in hydrostatic balance, and in
the scale invariant limit. This is described by Eq.~\ref{eq:6}, with
the dispersion relation Eq.~\ref{eq:5}, expressed in isopycnal
coordinates. By integrating analytically the two remaining Dirac
deltas, we simplify the collision integral reducing it to a double
integral. The wave kinetic equation thus takes the following form
{\begin{equation}\label{eq:11}
\begin{aligned}
	&\qquad\quad \qquad\qquad \qquad\partial_t n_\bp = \mI(k,m;a,b):= \int_0^\infty dk_{1}dk_2 \; {\cal J}(k,k_1,k_2,m)\,,\\
     &  {\cal J}(k,k_1,k_2,m)= \frac{8\pi}{k}\left(R^\bp_{12} f^\bp_{12} - R^1_{\bp2}f^1_{\bp2} - R^2_{\bp1}
        f^2_{\bp1}\right)\,, \quad R^\bp_{12} = k k_1 k_2
        |V^\bp_{12}|^2/\left(|{g^\bp_{12}}'|\Delta_{\bp12}\right)\,.
\end{aligned}
\end{equation}}
Here $ f^\bp_{12} = n_1n_2 - n_\bp (n_1 + n_2)\,$
and the area of the triangle of sides $k,k_1,k_2$, coming from
integration over angles under the assumption of isotropy, is given by
\begin{equation}
	\Delta_{\bp12} = \frac12 \sqrt{2(k^2k_1^2+k_1^2k_2^2+k^2k_2^2)-k^4-k_1^4-k_2^4}\,.\label{Delta}
\end{equation}
The expression of the matrix elements reads [\cite{iwthLPTN}]:
\begin{equation}\label{eq:12}
\begin{aligned}
	&V^\bp_{\bp_1 \bp_2} = \sqrt{kk_1k_2}\left(
  \frac{k^2+k_1^2-k_2^2}{2kk_1}\sqrt{\left|\frac{m_2^\star}{mm_1^\star}\right|}
  +
  \frac{k^2+k_2^2-k_1^2}{2kk_2}\sqrt{\left|\frac{m_1^\star}{mm_2^\star}\right|}
  + \frac{k^2-k_1^2-k_2^2}{2k_1k_2}\sqrt{\left|\frac{m}{m_1^\star
      m_2^\star}\right|} \right) \,,\\
	\end{aligned}
\end{equation}
\begin{equation}
	{g^\bp_{12}}' = \frac{\sign (m_1^\star)\; k_1}{(m_1^\star)^2}
        - \frac{\sign (m_2^\star)\; k_2}{(m_2^\star)^2} \,,
\end{equation}
where $m_1^\star, m_2^\star$ are given by the solution of the
resonance conditions, i.e. the joint conservation of momentum and
energy in each triadic interaction. Thus, in the four-dimensional
space spanned by $k_1$, $k_2$, $m_1$, $m_2$, the problem is now
restricted to the {\it resonant manifold}, parametrized by two
independent variables $k_1$ and $k_2$ as summarized in Table
\ref{tab:0}.
\begin{table}
\begin{center}
\begin{tabular}{ c |c |c }
Label $\;$&$\;$ Resonance condition$\;$ &$\;$ Solutions $\;$ \\ \hline
 $(\rm{Ia}),(\rm{Ib})$ & $\left\{\begin{array}{ll}
	\bp = \bp_1 + \bp_2\,\\
	\frac{k}{|m|} = \frac{k_1}{|m_1|} + \frac{k_2}{|m-m_1|}
\end{array}\right.$ & $\left\{\begin{aligned}
	&m_1^\star = \frac{m}{2k}\left[ k \pm k_1 \pm k_2 \pm \sqrt{(k\pm k_1 \pm k_2)^2 \mp 4 kk_1} \right]\\
	&m_2^\star = m - m_1^\star
\end{aligned}\right.$ \\  \hline
 $(\rm{IIa}),(\rm{IIb})$ & $\left\{\begin{array}{ll}
	\bp_1 = \bp + \bp_2\,\\
	\frac{k_1}{|m_1|} = \frac{k}{|m|} + \frac{k_2}{|m_1-m|}
\end{array}\right.$   & $ \left\{\begin{aligned} &m_2^\star
        = -\frac{m}{2k}\left[ k \mp k_1 - k_2 + \sqrt{(k\mp k_1 - k_2)^2 +
            4 kk_2} \right]\\ &m_1^\star = m + m_2^\star
\end{aligned}\right.\ $\\\hline
 $(\rm{IIIa}),(\rm{IIIb})$ & $\left\{\begin{array}{ll}
	\bp_2 = \bp + \bp_1\,\\
	\frac{k_2}{|m_2|} = \frac{k}{|m|} + \frac{k_1}{|m_2-m|}
\end{array}\right.$  &  $\left\{\begin{aligned}
	&m_1^\star = -\frac{m}{2k}\left[ k - k_1 \mp k_2 + \sqrt{(k-k_1 \mp k_2)^2 + 4 kk_1} \right]\\
	&m_2^\star = m + m_1^\star
\end{aligned}\right. $
\end{tabular}
\caption{The six independent solutions to the resonance conditions, [\cite{iwthLPTN})]. \label{tab:0}}
\end{center}
\end{table}
Note the symmetries of the resonant manifold: the solution $(\rm Ia)$
is obtained from solution $(\rm Ib)$ through permutation of the
indices $1\leftrightarrow 2$. We also notice that solutions $(\rm
IIa)$, $(\rm IIb)$ reduce to solutions $(\rm IIIa)$, $(\rm IIIb)$,
respectively, under permutation of the indices $1\leftrightarrow2$.

Based on the condition $b=0$ in Eq.~\eqref{eq:9}, we consider a
scale invariant solution {which is} horizontally isotropic and independent of the
vertical wave number:
\begin{equation}
	n_\bp = n(\bp) \propto k^{-a}\,.
\end{equation}
Since the collision integral is scale invariant in $k$, it is
sufficient to calculate it for a fixed value (e.g. $k=1$ for
simplicity), and then retrieve the solution for any value of $k$ by
using the scale invariance relation involving the homogeneity degree
of the collision integral.

\begin{figure}
\begin{center}
  \includegraphics[width=0.55\linewidth]{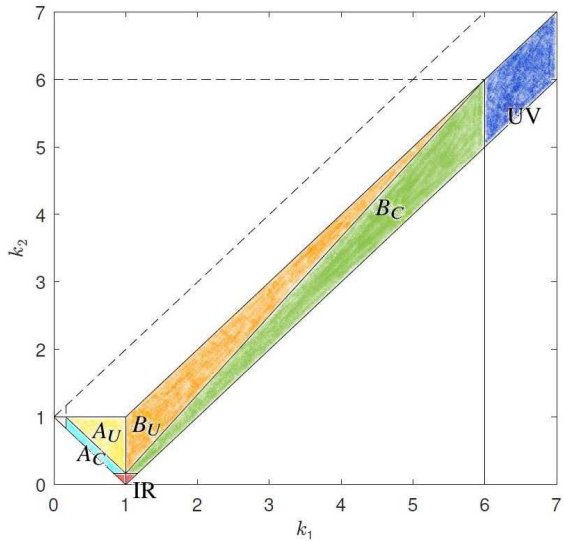}
\end{center}
  \caption{The kinematic box is splitted into subregions: $A_C$ (light
    blue), $A_{U}$ (yellow), $B_C$ (green), $B_{U}$ (orange), $\rm IR$
    (red), $\rm UV$ (blue). $A_C$ and $B_C$ are the near-colinear regions,
    $\rm IR$ and $\rm UV$ are the extreme scale-separated regions, and $A_{U}$
    and $B_{U}$ are the unclassified regions. A suitable
    Zakharov-Kraichnan transformation, see Eq.~\eqref{eq:40}, maps the
    regions $B_C$ and $B_{U}$ into $A_C$ and $A_{U}$, respectively.
 \label{fig:1}}
\end{figure}

Integration is performed in the kinematic box, defined by the three
triangular relations: $k+k_1\ge k_2$, $k_1+k_2\ge k$, $k+k_2\ge
k_1$. We differentiate three different regions of the kinematic box:
near-colinear region ($A_C$ and $B_C$), extreme-scale-separated region
(the infrared region $\rm IR$ and the ultaviolet region $\rm UV$), and the
region of unclassified triads, denoted as ($A_{U}$ and
$B_{U}$), as shown in Fig.~\ref{fig:1}.  Here, $A_C$ and $B_C$ are
named near-colinear regions since the resonant triads tend to the
colinear limit approaching their boundary given by $k_2=|k_1-k|$, a
relationship that can be fulfilled only by degenerate triangles with
their sides lying on the same line. The thickness of the regions $A_C$
and $B_C$ is given by the parameter $k_{\rm IR}$: small values of $k_{\rm IR}$ imply
that the resonant triads inside these regions are close to the
colinear limit.  We refer to $\rm IR$ and $\rm UV$ as the extreme-scale-separated
regions: for the triads in $\rm IR$, two wave numbers have finite
horizontal momentum, and one wave number has vanishing horizontal
momentum. For the triads in $\rm UV$, two wave numbers have very large
horizontal momentum, and one wave number has a much smaller horizontal
momentum. All the possible resonances in $\rm IR$ and $\rm UV$ constitute the
so-called named triads. Finally, $A_{U}$ and $B_{U}$ include all the
non-colinear, unclassified triads.
Exploiting symmetries, the r.h.s. of Eq.~\eqref{eq:6}, which we denote
by $\mI(k,m;a,b)$ after introducing the ansatz~\eqref{eq:9}, can be
reorganized as follows:
\begin{equation}\label{eq:10a}
\mI(k,m;a,b) = \Bigg[\int_{A_C} + \int_{A_{U}} + 2 \left(
\int_{B_C} + \int_{B_{U}} + \int_{\rm IR} + \int_{\rm UV} \right)\Bigg] \; {\cal J}(k,k_1,k_2,m) \;
dk_1dk_2\,,
\end{equation}
where a sum over the six solutions to the above resonance conditions
is implicit.  With $b=0$, the conditions on the exponent $a$ for convergence of
the collision integral on the r.h.s. of Eq.~\eqref{eq:11} come from
the infrared ($\rm IR$, red in Fig.~\ref{fig:1}) and the ultraviolet ($\rm UV$,
dark blue in Fig.~\ref{fig:1}) regions of integration. The details for
the computation of the following results are given in {\it
  Supplementary materials}~[\cite{GiovanniSupplemental}]. Both
singularities involve a first and a second cancellations between equal
and oppositely signed leading tems. For the infrared contribution, we
obtain
\begin{equation}\label{eq:30}
\begin{aligned}
	\mI_{\rm IR} & \simeq -16\pi a k^{-2a+4} m \int_0^{k_{\rm IR}/k} dx \int_{-x}^x
        dy\; x^{-a-1} \frac{y^2(y^2-x^2)}{\sqrt{x^2 - y^2}}\ =
        2\pi^2 \frac{a}{4-a}m k^{-a }k_{\rm IR}^{-a+4}\,,
\end{aligned}
\end{equation}
where $k_{\rm IR}$ is the (small) height of the red region in
Fig.~\ref{fig:1}. The integral converges if $a<4$. Also
    notice that the integral is {\it positive}. For the ultraviolet
contribution, we obtain
\begin{equation}\label{eq:39}
\begin{aligned}
	\mI_{\rm UV} &\simeq -32\pi a k^{-2a + 4}m\int_0^{k/k_{\rm UV}} dx \int_0^x dy\;
        \frac{k^2}{x^3} \;x^{a-8} \left[ (x-y)^4 + x^2(x-y)^2
          \right]/\sqrt{(2x-y)y}\\ &\simeq -14 \pi^2 \frac{a}{a-3}
        k^{-a+1} m\; k_{\rm UV}^{3-a}\, ,
\end{aligned}
\end{equation}
where $k_{\rm UV}$ is the $k_1$ coordinate of the left boundary of the
ultraviolet region. The integral converges when $a>3$. Note that this
contribution is {\it negative}, providing possibility for this
contribution to balance the positive contribution from
(\ref{eq:30}). This observation will later be exploited in Section
\ref{sec:2.3} to find the steady state solution composed of a balance
of infrared and ultraviolet contributions. The contribution
(\ref{eq:30}) is given by the resonance conditions $({\rm Ia})$ and
$({\rm IIa})$, the infrared ID resonances, while the contribution
(\ref{eq:30}) is given by the resonance conditions $({\rm IIb})$ and
$({\rm IIIc})$, the ultraviolet ID resonances. Both ES and PSI
resonances turn out to be subleading.

\subsection{Numerical solution: $a=3.69$}
Straightforward numerical integration can be performed only in $A_{\rm
  U}$ and $B_{\rm U}$, since close to the boundaries the integrand
contains integrable singularities. For this reason, numerical
integration is performed adopting the following technique for
integrable singularities [\cite{NC}]. We take the leading order
singularity of the integrand, integrate it analytically and add the
numerical integral of the difference between the integrand and the
leading order singularity. This way, the integrable singularities is
integrated analytically rather than numerically, ensuring accurate
results.  Notice that a singular behavior is found not only in the
infrared and ultraviolet regions, but also in the colinear regions,
due to the vanishing denominator $\Delta_{\bp12}$. The vanishing of
denominator occurs because the area of a triangle with colinear sides
tend to zero. The detail of the procedure for each of the five regions
is found in {\it Supplementary materials}
~[\cite{GiovanniSupplemental}].

To convince the reader that the numerical integration is performed
accurately, we show that the discretized integral is independent of
the step size of the discretization grid for sufficiently fine
grid. This is demonstrated in {\it Supplementary materials}
~[\cite{GiovanniSupplemental}].
\begin{figure}
\centerline{\includegraphics[width=0.5\linewidth]{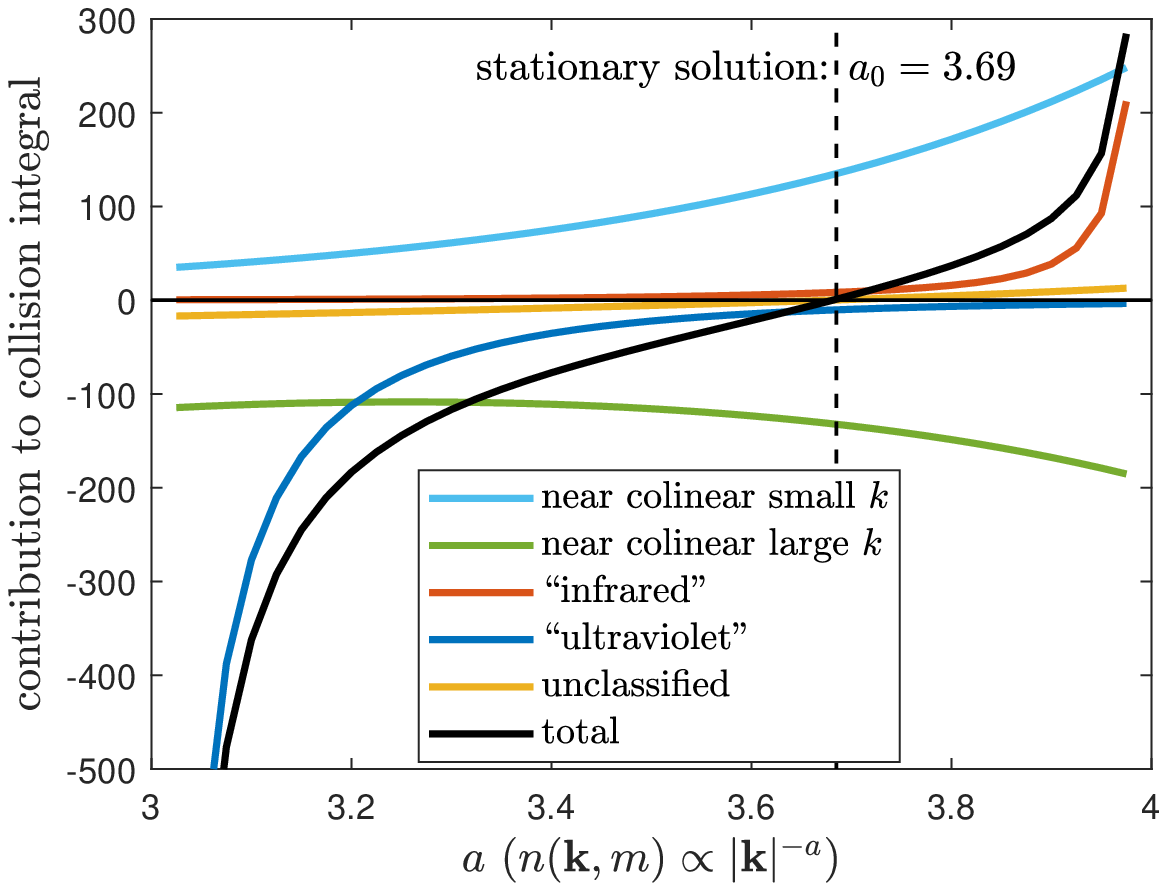}%
\includegraphics[width=0.5\linewidth]{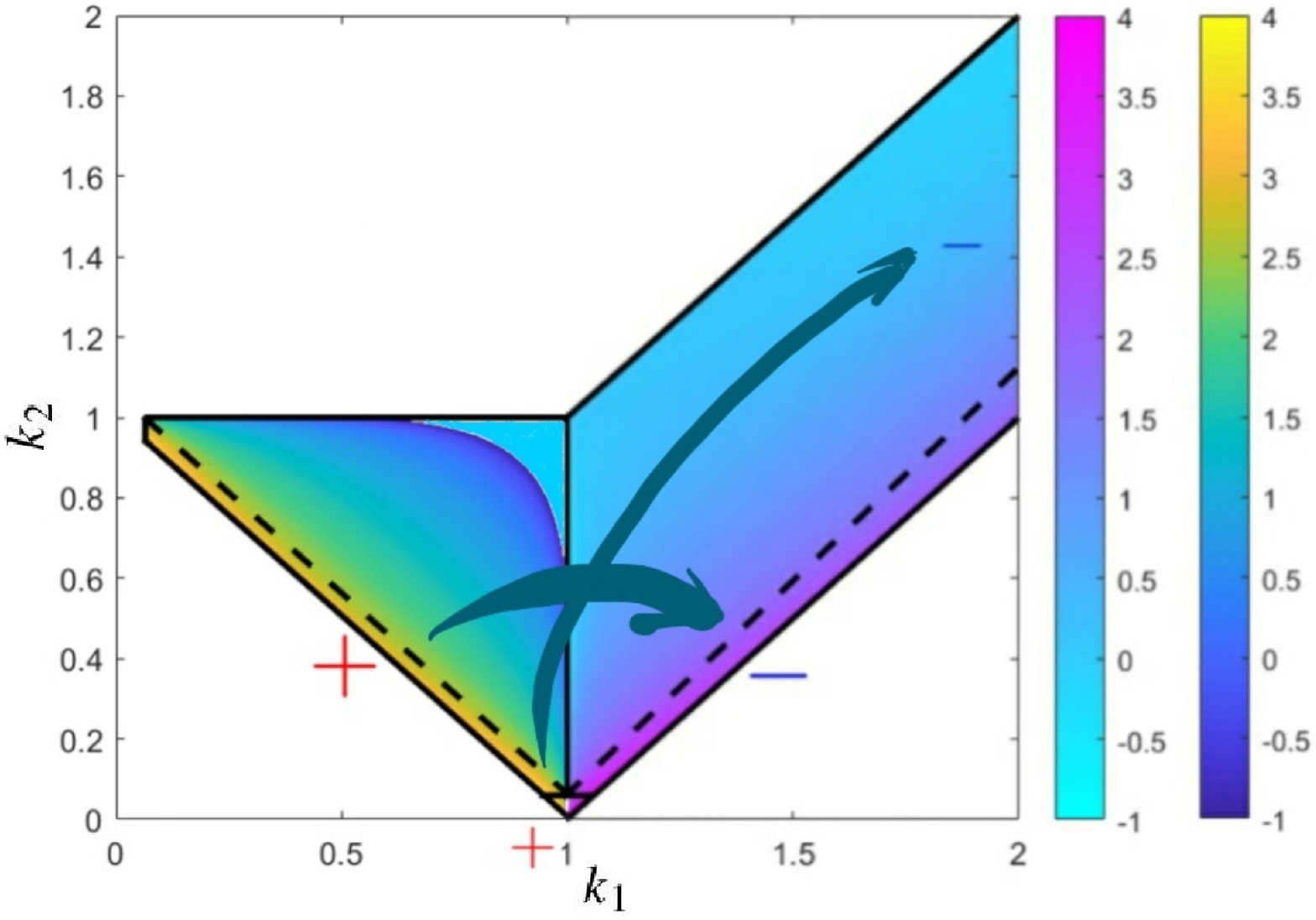}
}
  \caption{Left: Contributions of each subregion {(as split in Eq.~\eqref{eq:10a})} to the integral for
    $b=0$ and varying $a$. Right: The base-$10$ logarithm of the
    magnitude of the integrand is shown, for the solution $a=3.69$,
    $b=0$. The colormap labeled by the left colorbar indicates
    negative values, and the right colorbar indicates positive
    values. Also, we show here the schematic
    representation of the downscale energy transfers. The thicker
    arrow represents the stationary transfer between near-colinear
    regions, and the thinner arrow between regions with extreme scale
    separation. The fluxes of energy are explained in Section
    \ref{sec:3}.}
 \label{fig:6}
\end{figure}
The width of the regions around $k_2=0$ is determined by the parameter
$k_{\rm IR}$, while the cut at large $k$'s is performed at $k_1=k_{\rm UV}$. For the
result to be general, it must be independent of the choice of $k_{\rm IR}$
and $k_{\rm UV}$, as long as they are finite numbers, $k_{\rm IR}$ being
sufficiently small and $k_{\rm UV}$ sufficiently large. Iit turns out this is
indeed the case in our numerics. In {\it Supplementary materials}
~[\cite{GiovanniSupplemental}] we show how convergence is reached as
$k_{\rm UV}$ increases, as the neglected contribution in $\rm UV$
vanishes. Independence of the result upon variations of $k_{\rm IR}$ is even
more robust.

According to our results, the stationary spectrum of internal waves is given by
\begin{equation}\label{eq:3.69}
	n(k,m) \propto k^{-a_0}\,,\qquad a_0\simeq 3.69\,.
\end{equation}

The result in Fig.~\ref{fig:6}, {obtained with a choice
  $k_{\rm UV}=1/k_{\rm IR}=16$,} confirms convergence of the collision integral for
$3<a<4$: divergence of the integral is found both as $a\to3^+$ and
$a\to4^-$. In Fig.~\ref{fig:6}, we show the contributions of each
region of the kinematic box. The $a=3$ divergence is negative and due
to the region $\rm UV$ (ultraviolet). The contribution of $\rm UV$ is always
negative and tends to zero as $a\to4$. On the other hand, the $a=4$
divergence is positive and due to the region $\rm IR$. The contribution of
$\rm IR$ is always positive and tends to zero as $a\to3$. At the stationary
solution $a=a_0$ the contributions of $\rm IR$, $A_{U}$, $B_{U}$, $\rm UV$ are
close to zero, while the contributions of $A_{C}$ (positive) and
$B_{C}$ (negative) are large and cancel out. Notice that the results
are obtained for $A_C$ and $B_C$ being thin slices of width $1/16$:
the points in the colinear region correspond to triads of wave vectors
that have angles between each other's horizontal components of $3^\circ$ or less!

Let us consider an arbitrary reference number $k=1$.  The sign of the
integrand in the right panel of Fig.~\ref{fig:6} indicates the
direction of the energy transfers. We see from Fig.~\ref{fig:6} that
the contribution from the wave numbers with $k_1<1$ is
positive. Therefore there is a net energy flow from wave numbers
smaller than reference number $k=1$ to the reference wave number
$k=1$. On the other hand, the contribution from the wave numbers with
$k_1>1$ is negative. Consequently the wave number $k=1$ constantly
pumps energy towards higher wave numbers. Therefore we conclude that
the energy transfer is directed towards high horizontal wave numbers.
We elaborate on this further in Section \ref{sec:3}.\\

The outflowing energy from the small-wave-number near-colinear triads
is balanced by the inflowing energy at the large-wave-number
near-colinear triads, implying a stationary flow mediated by $k=1$
which is represented as a thick arrow in Fig.~\ref{fig:6}. The
outflowing energy from the infrared region is balanced by the
inflowing energy entering the ultraviolet region,
giving a stationary energy flow between these two regions mediated by
$k=1$. This energy transfer is represented as a thin directed arrow
connecting the two regions. The quantitative justification of the
balance is given by the left panel of Fig.~\ref{fig:6}.

The value of the exponent appears to be characterized importantly by a
balance of the regions $A_C$ and $B_C$. This suggests that a suitable
transformation mapping one region into the other could potentially
make the search for the steady self-similar spectrum amenable to
analytical treatment. This task is addressed in the next section.
\section {Analysis of the contributions to the stationary solution}\label{sec:2}
\noindent We consider the reorganized expression of the collision
integral in Eq.~\eqref{eq:11}. Let us introduce the following
Zakharov-Kraichnan transformations,
\begin{equation}\label{eq:40}
\left\{ \begin{aligned}
& k_1 = \frac{k^2}{\tilde k_1} \ 
& k_2 = \frac{k \tilde k_2}{\tilde k_1}
\end{aligned}\right.\,,  \qquad\qquad\qquad
\qquad\left\{ \begin{aligned}
& k_2 = \frac{k^2}{\tilde k_2}\ 
& k_1 = \frac{k \tilde k_1}{\tilde k_2}
\end{aligned}\right.\,,
\end{equation}
and notice that under the first transformation the regions $B_C$ and $B_{U}$ are
mapped into $A_C$ and $A_{U}$, respectively, if the choice
$k_{\rm UV}=k^2/k_{\rm IR}$ is made. In the following we consider the contributions
from the three types of regions (near-colinear,
extreme-scale-separated and unclassified) separately, with the goal of
locating the regions and the resonances that matter the most and the
ones that are negligible.

\subsection {Colinear limit}\label{sec:2.1}
\begin{figure}
\includegraphics[width=\linewidth]{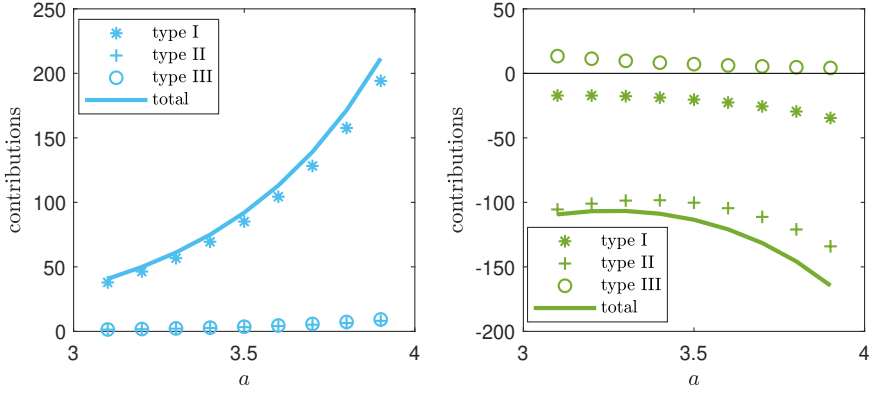}
  \caption{The contributions from $A_C$, left panel, and $B_C$, right
    panel (the latter multiplied by $2$ to account for its symmetric
    $B_C'$ by permutation $k_1\leftrightarrow k_2$), are splitted into
    their sub-contributions from the three resonance types, showing
    that $A_C$ is dominated by the contribution $({\rm I})$, while
    $B_C$ is dominated by the contribution $({\rm II})$. Computing the
    balance between these two contributions leads to a theoretical
    estimate $a = 7/2$.
 \label{fig:7}}
\end{figure}

\noindent We start by analyzing the contributions in the regions $A_C$
and $B_C$ by decomposing them into separate sub-contributions from the
three resonance types $({\rm I})$, $({\rm II})$ and $({\rm III})$ (see
Table \ref{tab:0}). The different contributions are plotted in
Fig.~\ref{fig:7}. We realize that in the region $A_C$ the leading
contribution is given by the resonance condition $({\rm I})$, while in
region $B_C$ the leading contribution is given by the resonance
condition $({\rm II})$. Let us consider only these two contributions,
naming them the {\it main colinear contributions}:
\begin{equation}
\begin{aligned}
	\mI_{A_C}+2\mI_{B_C} \simeq & \int_{k_{\rm IR}}^{1-k_{\rm IR}} dk_2
        \int_0^{k_{\rm IR}} dx \frac{T^k_{k_1, k-k_1}}{\sqrt{2kk_1(k-k_1)x}}
        \\ & - \int_{1+k_{\rm IR}}^{k_{\rm UV}} dk_2 \int_0^{k_{\rm IR}} dx
        \frac{T^{k_1}_{k, k_1-k}}{\sqrt{2k k_1(k_1-k)x}} -
        \int_{1+k_{\rm IR}}^{k_{\rm UV}} dk_2 \int_0^{k_{\rm IR}} dx \frac{T^{k_2}_{k,
            k_2-k}}{\sqrt{2k k_2(k_2-k)x}}\,,
\end{aligned}
\end{equation}
where $T^0_{12} = kk_1k_2 |V^0_{12}|^2 f^0_{12}/|{g^0_{12}}'|\,.$
Transforming the integral into an integral in the region $A_C$, by
using the symmetries of Eq. \eqref{eq:40} (with
$k_{\rm UV}=k^2/k_{\rm IR}$) and the scale-invariant properties of the integrand, we
obtain (renaming $\tilde k_1 \rightarrow k_1$, $\tilde k_2 \rightarrow
k_2$)
\begin{equation}\label{eq:49}
	\mI_{A_C}+2\mI_{B_C} \simeq  \int_0^{k_{\rm IR}} dx \int_{k_{\rm IR}}^{1-k_{\rm IR}} dk_2  \frac{T^k_{k_1, k-k_1}}{\sqrt{2kk_1(k-k_1)x}} \left[ 1 - \left( \frac{k}{k_1} \right)^{r+3} - \left( \frac{k}{k_2} \right)^{r+3} \right]\,,
\end{equation}
where $r$ is the degree of homogeneity of the integrand: $r=3-2a$
(same dependence on $k$ as in the computation of the
Pelinovski-Raevski spectrum). We used the property of the
Zakharov-Kraichnan transformation for an homogeneous function of
degree $r$ with respect to horizontal wave numbers:
\begin{equation}
	\mathcal{J}\left(k_1,k,k_2,m\right)=\mathcal{J}\left(\frac{k}{\tilde k_1}k,
        \frac{k}{\tilde k_1} \tilde k_1,\frac{k}{\tilde k_1} \tilde
        k_2,m\right) = \left(\frac{k}{\tilde k_1}\right)^r \mathcal{J}\left(\tilde
        k,\tilde k_1,\tilde k_2,m\right)\,,
\end{equation}
and a factor $(k/\tilde {k_1})^3$ appeared due to the Jacobian of the
coordinate change. Now, we notice that if $r+3=-1$, the integrand
contains a factor $[k-k_1-k_2]$. Since we are considering the
near-colinear region, {horizontal} momentum conservation implies that such a factor
vanishes (more precisely, it would vanish in the limit $k_{\rm IR}\to0$):
$A_C$ is a thin slice lying on the line $k_2 = k-k_1$. Therefore, we
have that $\mI_{A_C}+2\mI_{B_C}=0$ for $6-2a=-1$, which determines the
value of $a=7/2$ as the critical exponent\,.

We have therefore found analytically the steady state solution for the
reduced kinetic equation dominated by the balance between the
contribution of type $({\rm I})$ in the region $A_C$ and the
contribution of type $({\rm II})$ in the region $B_C$, which are the
largest contributions from the near-colinear regions. This solution is
therefore given by
\begin{equation}\label{ThreePointFive}
n(k,m) =
  k^{-\frac{7}{2}}m^0\,.
\end{equation}
Since this solution is close to the Pelinovsky-Raevsky spectrum
(\ref{PelinovskyRaevsky}), we propose to call this solution
{\it modified  Pelinovsky-Raevsky spectrum}.
The difference between~\eqref{ThreePointFive}
and~\eqref{PelinovskyRaevsky} is that the latter is the {\it formal}
solution, {corresponding to a {\it non-local} spectrum (i.e., implying a divergent collision integral). The former solution, on the other hand, is a physically
relevant solution corresponding to a {\it local} action spectrum (i.e., whose collision integral is finite)}. Note, however that a part of the resonances have been
neglected.

The result $a=7/2$ coming from the colinear limit of Eq. \eqref{eq:49}
involves only the two leading contributions in Fig.~\ref{fig:7}. This
observation provides an intuition on how the exponent $a$ is
determined by the kinetic equation.  The sum of the subdominant
contributions in Fig.~\ref{fig:7} is negative and almost independent
of $a$. When added to the main contribution crossing zero at $a=7/2$,
this negative contribution makes the zero-crossing point shift toward
the right and in Fig.~\ref{fig:11a}{\bf (A)} the total colinear contribution is
shown to cross zero at $a\simeq 3.69$.
\subsection {Extreme-scale-separated triads}\label{sec:2.3}
\begin{figure}
  \begin{tabular}{cc} {\bf{A}}& {\bf{B}}\\
\includegraphics[width=0.5\linewidth]{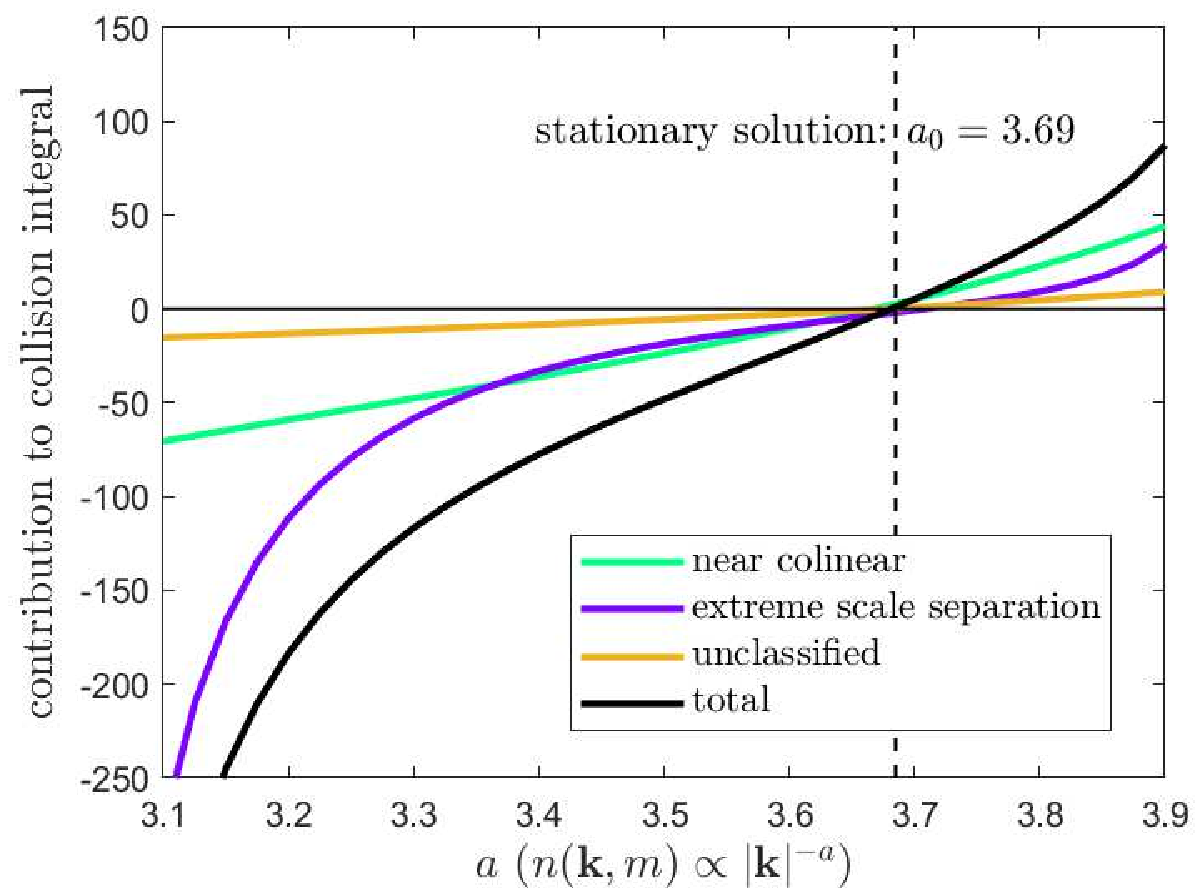}
  &
\includegraphics[width=0.5\linewidth]{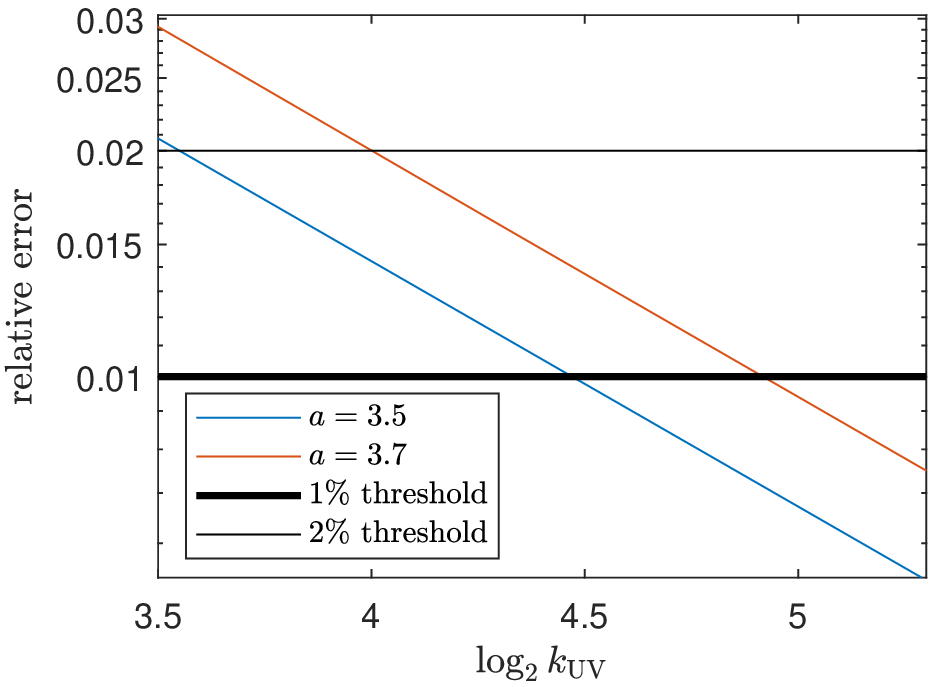}
\end{tabular}
  \caption{{\bf A} The net contribution of the three types of triads, showing
    that each type balances to zero independently at the convergent
    stationary solution. {\bf B}
    Relative error of Eq.~\eqref{eq:39} with respect to the
    fully numerically integrated contribution of region ${\rm UV}$. Around
    $a=3.7$, we see that a choice of $k_{\rm UV}=16$ ($y=4$) implies an error
    of about $2\%$, which we consider acceptable. Thus, we consider
    $k_{\rm UV}=16$ as a reasonable delimitation of the ultraviolet region,
    as well as $k_{\rm IR}=1/16$ as the delimitation of the infrared region.}
 \label{fig:11a}
\end{figure}

\noindent In this section we consider the contribution that comes from
the extreme-scale-separated triads, the sum of the infrared and the
ultraviolet contributions. The former is positive and tends to
$+\infty$ as $a\to4^-$; the latter is negative and tends to $-\infty$
as $a\to3^+$. In Fig.~\ref{fig:11a}{\bf (A)} we show the total contribution of
the extreme-scale-separated resonances and we observe numerically that
it crosses zero around $a\simeq 3.69$, too.
In Section~\ref{sec:loccond} we proposed a balance between positive
infrared~\eqref{eq:30} and negative ultraviolet~\eqref{eq:39}
contributions as a way to form the steady state spectrum of internal
waves.  This balance hinges upon the choice $k_{\rm UV}=k^2/k_{\rm IR}$, as
explained in Section~\ref{sec:2.1}.

In Fig.~\ref{fig:11a}, we show the relative error of the leading order
analytical expression~\eqref{eq:39}, with respect to the numerically
computed contribution, as a function of $k_{\rm UV}$. The result is shown for
$a=3.5$ and $a=3.7$. Using the expression $k_{\rm UV}=2^y$, with $k=1$, we
observe that the analytical approximation is good starting from values
of $y$ between $4$ and $5$. We choose $y$ (and therefore $k_{\rm UV}$ and
$k_{\rm IR}$) large enough for an accurate approximation with the leading
order expression~\eqref{eq:39} (or \eqref{eq:30}), yet small enough so
that all extreme-scale-separated triads (\eqref{eq:39}) and
(\eqref{eq:30}) are actually included in the ultraviolet and infrared
regions.  We make an arbitrary choice $y=4$ ($k_{\rm UV}=1/k_{\rm IR}=16$) so that
the leading order error of (\eqref{eq:30}) and (\eqref{eq:39}) is
about 2 percent (see Fig.~\ref{fig:11a}). Our results are insensitive
to this specific choice. The balance between the two expressions gives
\begin{equation}\label{eq:51}
\begin{aligned}
	2\pi^2 \frac{a}{4-a} k_{\rm IR}^{4-a} &= 14\pi^2
        \frac{a}{a-3}\frac{1}{k_{\rm IR}^{3-a}}\\ h_y(a):=(2a-7) y \log 2 &=
        \log 7 +\log\left(\frac{4-a}{a-3}\right)=:g(a)\,.
\end{aligned}
\end{equation}
First, we notice that the presence of a factor $7$ on the
r.h.s. breaks the symmetry that would imply the two contributions to
balance out at $a=7/2$, in the middle of the convergence interval
$(3,4)$. Secondly, we notice that the function $g(a)$ has an
inflection point at $a=7/2$, making its Taylor expansion of first
order have an error of third order. Using linear interpolation
centered at $a=7/2$, $g_{7/2}(a)=\log7 -2(2a-7)$, as an accurate
approximation to $g(a)$, demanding that $h_y(a) = g_{7/2}(a)$ we
obtain
\begin{equation}
	a= \frac72 +\frac{\log 7}{2y\log2 + 4}\,.
\end{equation}
If we adopt $y=4$, as chosen throughout the paper, we obtain the
solution $a\simeq3.70$. Using $y=5$ would yield a solution
$a\simeq3.68$.
 
In this section, we have shown how the formation of the
  stationary solution of the wave kinetic equation can be interpreted
  via two independent balances, between near-colinear triads and
  between the ID triads of the extreme-scale-separated regions. This
  is consistent with recent results from direct numerical simulations
  where the dominating interactions were located in the ID regime but
  also throughout all of the interval $k_1\in[0,1.4k]$
  [\cite{pan2020numerical}]. Despite the latter effect was there
  investigated as a broadening of ID due to large nonlinearity, we
  find that it may be consistent with the colinear resonances depicted in the
  right panel of Fig.~\ref{fig:6}.

\section{Downscale energy transfers}\label{sec:3}
\subsection{Physical dimensions and energy conservation}\label{sec:3.1}

\noindent Using the scale-invariant properties of the collision
integral, the r.h.s. of Eq.~\eqref{eq:11} can be rewritten considering
the appropriate physical dimensions as
\begin{equation}\label{eq:63}
	I(\bk,m) = |m|^{-2b+1} k^{-2a+4} (V_0A)^2 \mI(k=1,m=1;a,b)\,,
\end{equation}
where $\mI(k=1,m=1;a,b)$ is non-dimensional, $V_0$ is the dimensional
prefactor of the matrix element defined below in Eq.  (\ref{eq:64}),
and $A$ is the prefactor of the Garrett and Munk spectrum defined in
(\ref{eq:1b}) above.  A simple way to check the dimensional
consistency of the prefactor in~\eqref{eq:63} is to consider the
contribution from the extreme-scale-separated region,
Eq.~\eqref{eq:51}, which is analytically tractable:
\begin{equation}
	2\pi^2 \frac{a}{4-a} k^{-a} k_{\rm IR}^{4-a} - 14\pi^2
        \frac{a}{a-3}\frac{k^{-a+1}}{k_{\rm IR}^{3-a}} = mk^{-2a+4} \left(
        2\pi^2 \frac{a}{4-a} x_{\rm IR}^{4-a} - 14\pi^2
        \frac{a}{a-3}x_{\rm IR}^{a-3} \right)\,,
\end{equation}
where $x_{\rm IR}=k_{\rm IR}/k$ and the term in brackets is a nondimensional
function of the exponent $a$ that vanishes at $a=a_0$. The factor
$(V_0 A)^2$ comes from having both the matrix elements and the
spectrum to the second power in the collision integral, and by
introducing the appropriate dimensional constants that have been
omitted so far.  The dimensional properties of the collision integral
are indeed the same also in the other integration regions.

Next, we compute the spectral energy fluxes, recalling that
the energy density is given by
\begin{equation}
	e(k,m) = 2\pi k \sigma(\bk,m) n(\bk,m)\,, \qquad
        \sigma(\bk,m)=\alpha \frac{k}{|m|}\,,\;\text{ with
        }\;\;\alpha=\frac{g}{\rho_0 N}\,,
\end{equation}
where the scale invariant dispersion relation~\eqref{eq:5} is
used. Using~\eqref{eq:63}, the stationary wave kinetic equation for
the energy density assumes the simple form
\begin{equation}\label{eq:100}
\dot e(k,m) = 2\pi (A V_0)^2 k^{-2a+6} m^{-2b} \mI(k=1,m=1;a,b)=0\,.
\end{equation}

\subsection{Dimensional prefactors}\label{sec:3.3}
\noindent The dimensional factor coming from the matrix elements in
Eq.~\eqref{eq:63} is given by
\begin{equation}\label{eq:64}
	|V_0|^2= \frac{N}{\color{black}32 \rho_0}\,.
\end{equation}
The factor $A$ is the dimensional prefactor of the GM spectrum, our
observational input for the oceanic wave field. The procedure to
obtain $A$ is explained in the rest of this paragraph. The
{non-rotating limit} of the GM spectrum in $k-m$ coordinates is given
by Eq.~\eqref{eq:4}. However, in isopycnal coordinates the spectrum
needs to be multiplied by a factor $N^2\rho_0/g$, which gives
\begin{equation}\label{eq:GM}
	n_{\rm GM}(k,m) = \frac{{\color{black}1}}{\pi^3 g} E b^2 {\color{black}N_0N} \rho_0 f m_\star
        k^{-4}\,, \qquad m_\star= 3\pi\frac{N}{b N_0}\,,
\end{equation}
where $E=6.3\times 10^{-5}$ is the non-dimensional {\it energy level},
$b=1300$ m, $\rho_0=1000$ kg$/$m$^3$, $N_0=0.00524$ s$^{-1}$, $f =
2\cdot 7.3\times 10^{-5} \sin(l)$ (at latitude $ l=32.5^\circ$){\footnote{A pre-factor of $4$ instead of $3$ is known to imply a more accurate asymptotic fit in the large wave number regime [\cite{regional}]. However, for simplicity here we keep the factor appearing in the original 1976 GM parametrization, as is. We will see that the choice does not affect the order of magnitude of the estimate of the flux.}}. The
values given here are the ones of the standard GM
parametrization. Since for the GM spectrum we have $a=4$ instead of
$a=3.69$, we introduce the modified version of Eq.~\eqref{eq:GM}:
\begin{equation}\label{eq:GMc}
	n_{\rm GM,c}(k,m) = \frac{{\color{black}1}}{\pi^3 g} E b^2 {\color{black}N_0N} \rho_0 f
        \frac{m_\star}{{k_\star}^{(1-s)/2}} k^{-(4-(1-s)/2)}\,, \quad
        k_\star= m_\star/r\,,
\end{equation}
where $ s=2a-7 = 0.38$. This is a slightly less steep, dimensionally
consistent version. The non-dimensional parameter $r={\color{black}N}/f$ quantifies
the horizontal-to-vertical anisotropy, [\cite{regional}].  Now, the GM
spectrum is normalized so that the total energy is expressed in units
of J$/$kg, as usual in physical oceanography. On the other hand, in
the wave turbulence formalism the total energy is expressed as a
density per unit of volume of the physical space. Here, the physical
space has units of ${\rm kg}/{\rm m}$, given by an area in the
horizontal directions times a density in the vertical. It is therefore
possible to switch from one representation to the other multiplying by
the appropriate density, which in this case is the characteristic
density of the isopycnal coordinates, ${\color{black}\Pi=g/N}^2$, in units of
meters (normalized differential thickness of an isopycnal layer). This
results into the equivalence: $n_{\rm WKE}=n_{\rm GM} \cdot g/{\color{black}N}^2$, which
applied to~\eqref{eq:GMc} finally gives the dimensionally consistent
factor
\begin{equation}\label{eq:A}
 A=\frac{{\color{black}1}}{\pi^3} E b^2 \rho_0 f {\color{black}\frac{N_0}{N}}\frac{m_\star}{k_\star^{(1-s)/2}}\,.
\end{equation}

\subsection{Dissipated power at high wave numbers}
In order to compute the energy flux towards high wave numbers we need
to consider the physical cutoffs of the problem. Natural cutoffs are
imposed on the vertical wave number by the depth of the ocean and by
the wave breaking cut off, and on the frequency by the inertial
frequency and the buoyancy frequency,
\begin{equation}
	m_{\rm min} =\frac{2\pi g}{\rho_0 N^2}\times(2600 \text{
          m})^{-1}\,\quad m_{\rm max} =\frac{2\pi g}{\rho_0
          N^2}\times(10 \text{ m})^{-1}\,,\quad \sigma_{\rm min} =
        f\,,\quad \sigma_{\rm max} = N\, .
\end{equation}
These limiting values define a rectangle in $\sigma - m$ space that
translates into a trapezoid in $k-m$ space, with inclined sides given
by
\begin{equation}\label{eq:kmax}
	k_{\rm min}(m) =\frac{f}{\alpha}m\,,\quad k_{\rm max}(m)
        =\frac{N}{\alpha}m\,.
\end{equation}
The collision integral contains all of the necessary information on
the spectral energy transfers.  In the following, we compute
numerically the outflowing power at high wave numbers from computation
of the contribution of the resonant triads with an output wave number
such that $m>m_{\max}$ or $k>k_{\max}(m)$, i.e. assuming that the
production of a wave beyond the physical high wavenumber cutoff
results into complete dissipation of its energy. At the same time, it is assumed that the region within the physical cut-offs is an {\it inertial range} with no sources nor sinks, where energy is transfered exclusively via resonant interactions.

Let us define the part of the collision integral which contributes to
the dissipation of energy by transferring it beyond the dissipation
cut-offs:
\begin{equation}\label{eq:5.11}
\begin{aligned}
	I_{\rm diss}(\bk,m;k_{\max}) :&
	 =  \frac{{{\color{black}N}}^2}{g}  (V_0 A)^2 \alpha^{-1} |m|^{-2b+1} k^{-2a+4} \mI_{\rm diss}(\kappa)\,,\\
\mI_{\rm diss}(\kappa) &= \int_{\Omega_{\rm h}(\kappa)} \mathcal J(k=1,k_1,k_2,m=1) dk_1 dk_2\,,
\end{aligned}
\end{equation}
where $\kappa = k_{\max}/k$ and $\Omega_{\rm h}(\kappa)$ as the set of
triads transferring energy to output waves beyond the horizontal
dissipation cutoff $k_{\max}$, and is defined in Table
\ref{tab:1}. Moreover, a factor $\frac{{{\color{black}N}}^2}{g}$ is added to
account for transition to isopycnal coordinates, the inverse of the
factor in Eq. \eqref{eq:A}.  Here, $I_{\rm diss}(\bk,m;k_{\max})$
quantifies the amount of wave action that wave number $\bp$ sends via
resonant interactions beyond the dissipation threshold $k_{\max}$, per
unit time and per unit volume of {\color{black}Fourier} space.
\begin{table}
\begin{center}
\begin{tabular}{ c |c |c |c}
Input $\;$&$\;$ Output$\;$ &$\;$ $\Omega_{\rm h}(\kappa)$$\;$ &$\;$ $\Omega_{\rm v}(\mu)$ \\ \hline
 $\bp$ & $\bp_1,\bp_2$ & $\left. \begin{array}{l} k_1>\kappa,k_2>\kappa\\  k_1>\kappa,k_2<\kappa: \text{ weight } \tfrac{\sigma_1}{\sigma_1+\sigma_2}\\  k_1<\kappa,k_2>\kappa: \text{ weight } \tfrac{\sigma_2}{\sigma_1+\sigma_2} \end{array} \right.$ & $\left. \begin{array}{l} |m_1|>\mu,|m_2|>\mu\\  |m_1|>\mu,|m_2|<\mu: \text{ weight } \tfrac{\sigma_1}{\sigma_1+\sigma_2} \\  |m_1|<\mu,|m_2|>\mu: \text{ weight } \tfrac{\sigma_2}{\sigma_1+\sigma_2} \end{array} \right.$ \\  \hline
 $\bp,\bp_2$ & $\bp_1$   &  $k_1>\kappa,k_2<\kappa $ & $|m_1|>\mu,|m_2|<\mu $ \\\hline
 $\bp, \bp_1$ & $\bp_2$  &  $k_1<\kappa,k_2>\kappa $ & $|m_1|<\mu,|m_2|>\mu $
\end{tabular}
\caption{All of the non-negligible contributions for $\kappa>1$ are
  negative (see right panels of Figs. \ref{fig:6}-\ref{fig:7}). The
  same holds for $\mu>1$. Thus, wave number $\bp$ is always an
  incoming wave in the triads here considered. The table represents
  the conditions under which every type of resonance results into the
  dissipation of energy due to at least one output beyond the
  dissipation cutoffs. During the numerical computation of the
  integrals $C_{\rm h}$ and $C_{\rm v}$, these conditions are applied
  for every point of the kinematic box. In the case of a decay into
  two waves of which only one is dissipated, only the fraction of
  energy of the dissipated wave must be accounted for, multiplying the
  contribution by the weight shown in the table.\label{tab:1}}
\end{center}
\end{table}
The power (per unit of {\color{black}$m$}) dissipated beyond
$k_{\max}$ at fixed $m$ {\color{black}(now considered as the positive definite magnitude  of $m$)} is related to the integral of $I_{\rm
  diss}(\bk,m)$ for all values of $k$:
\begin{equation}\label{eq:Fdiss}
\begin{aligned}
	F_{\rm diss}(m) &= \int_{k_{\min}}^{k_{\max}} {\color{black}4}\pi k \omega(\bp) I_{\rm diss}(\bk,m;k_{\max}) dk\,\\
	&= \frac{{{\color{black}N}}^2}{g}\int_{k_{\min}}^{k_{\max}} {\color{black}4}\pi |m|^{-2b+1}k^{-2a+5} \alpha \frac{k}{|m|} (V_0 A)^2 \alpha^{-1} \mI_{\rm diss}(\kappa)\\
	&= {\color{black}4}\pi \frac{{{\color{black}N}}^2}{g} (V_0 A)^2 k_{\max}^{-s} C_{\rm h}\,,\quad \text{with}\qquad C_{\rm h} := \int_1^{\frac{N}{f}} \kappa^{s-1} \mathcal{I_{\rm diss}(\kappa)}\,,
\end{aligned}
\end{equation}
where we recall that $s=2a-7=0.38$, and $b=0$.
In a similar fashion, we can obtain an analogous relation for the
energy flux dissipated vertically at vertical wave numbers larger than $m_{\max}$:
\begin{equation}\label{eq:Gdiss}
\begin{aligned}
	&G_{\rm diss}(k) = {\color{black}4}\pi \frac{{{\color{black}N}}^2}{g} (V_0 A)^2
        k^{-(1+s)}{m_{\max}} C_{\rm v}\,,\qquad C_{\rm v} :=
        \int_1^{\frac{m_{\max}}{m_{\min}}} \mu^{-2} \mathcal{K_{\rm
            diss}(\mu)}\,,\\
	&\qquad\qquad \mathcal K_{\rm diss}(\mu) = \int_{\Omega_{\rm v}(\kappa)} \mathcal J(k=1,k_1,k_2,m=1) dk_1 dk_2\,,
\end{aligned}
\end{equation}
where $\mathcal K_{\rm diss}(\mu)$ is the analogue of $\mI_{\rm
  diss}(\kappa)$ in the vertical direction, i.e. the collision
integral restricted to the triads with output waves beyond the
$m_{\max}$ threshold. The conditions defining $\Omega_{\rm v}$ are
found in Table \ref{tab:1}. These integrals are computed
numerically. {The computation of $C_{\rm h}$ is quite straightforward
  since the kinematic box is expressed in horizontal wave number
  coordinates. The quantity $\mI_{\rm diss}(\kappa)$ is computed
  inside a loop spanning all values of $\kappa$, checking the
  constraint of $\Omega_{\rm h}(\kappa)$ for every point. Note that since the
  position of the right boundary depends on $m$, \eqref{eq:kmax}, in
  the computation of $\mI_{\rm diss}(\kappa)$ we have to impose that
  $k_1>k_{\max}m_1/m$ (or the same for $k_2$) for point $(k_1,m_1)$ to
  be past the absorbing boundary.  For the computation of $C_{\rm v}$, on
  the other hand, at every loop iteration we
  have to integrate over all of the kinematic box to compute $\mathcal
  K_{\rm diss}$ and check point by point whether $m_1$ or $m_2$ have
  ``crossed'' the boundary at $m_{\max}$ (which is independent of
  $k$).

  One important point requires particular care. In the
  previous sections we relied upon the leading order of the infrared
  contribution of Eq.~\eqref{eq:30}, which is obtained after the
  second cancellation where a negative singularity for $k_1>1$ cancels
  exactly with a positive singularity for $k_1<1$. However, since now
  in Eq.~\eqref{eq:5.11} we consider an integration region where
  $k_1>\kappa>1$, the positive singularity due to $k_1<1$ is no longer
  present and thus the use of Eq.~\eqref{eq:30} is not justified. In the integrand of
Eq.~\eqref{eq:5.11}, whose expression is found in the {\it Supplementary materials}
[\cite{GiovanniSupplemental}], the finite point singularity therefore has exponent $-a+3/2$ rather than the $-a+2$ resulting after the second cancellation (leading to Eq.~\eqref{eq:30}). Integrating twice according to Eq.~\eqref{eq:5.11}, the exponent of the singularity of the integrand defining $C_{\rm h}$ in Eq.~\eqref{eq:Fdiss} is $\beta_{\rm h} = -a+7/2\simeq 0.19$\,. Numerical computation of $\mI_{\rm diss}(\kappa)$ as $\kappa\to1^+$ confirms this analytical prediction and gives a best least square fit of $\mI_{\rm
  diss}(\kappa) \sim -d_{\rm h}(\kappa-1)^{-\beta_{\rm h}}$, for $\kappa-1\ll 1$, with
$d_{\rm h}={\color{black}127}\times8\pi$, $\beta_{\rm
  h}=0.19$. An analogous computation for the integral $C_{\rm v}$ defined in Eq.~\eqref{eq:Gdiss} leads to a best fit scaling of its integrand of $\mathcal K_{\rm diss}(\mu) \sim -d_{\rm v}(\mu-1)^{-\beta_{\rm v}}$, for $\mu-1\ll1$,
with $d_{\rm v}\simeq9.0\times8\pi$, $\beta_{\rm v}\simeq0.75$. In this case, an analytical evidence of the exponent is not simply available since the kinematic box is not expressed in vertical wave number coordinates. In Fig.~\ref{fig:15}, we show the rapidly decaying behaviour of the integrands of $C_{\rm h}$ and $C_{\rm v}$. Therefore, $C_{\rm h}$ and $C_{\rm v}$ are ``universal constants'' which are relatively insensitive to the limits of integration in Eqs.~\eqref{eq:Fdiss}-\eqref{eq:Gdiss}. The figure also shows a closer look to the scaling of the singularities as $\kappa\to1^+$ and $\mu\to1^+$, with the  respective best fit scalings.
Therefore we obtain:
\begin{equation}\label{eq:5.16}
\begin{aligned}
	C_{\rm h} \simeq \int_1^{\frac{N}{f}} \kappa^{s-1}
        \mathcal{I_{\rm diss}(\kappa)} &\simeq -\frac{d_{\rm h}
          \,{k_{\rm IR}}^{1-\beta_{\rm h}}}{1-\beta_{\rm h}} +
        \int_{1+k_{\rm IR}}^{\frac{N}{f}} d\kappa \,\kappa^{s-1}
        \mathcal{I}_{\rm diss}(\kappa)\\ &\simeq -8\pi{\color{black}(16.6+54.6+10.2)
        \simeq -8\pi\times 81.4}\,.
\end{aligned}
\end{equation}
Here the three numbers {\color{black}$16.6$, $54.6$ and $10.2$} represent the partial
contributions by the infrared ID interactions, near-colinear
interactions and unclassified interactions, respectively. The
ultraviolet contribution is negligible.
\begin{figure}
\centerline{\includegraphics[width=\linewidth]{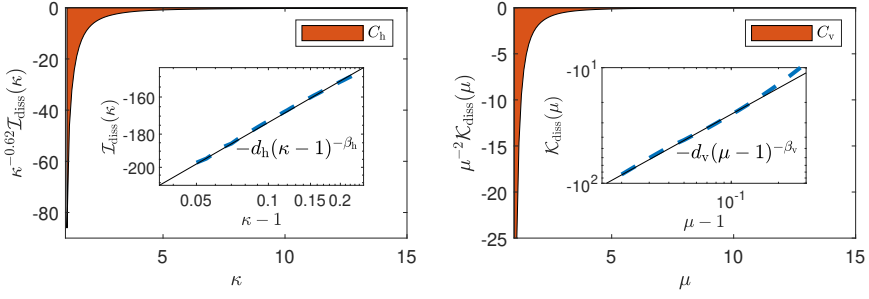}}
  \caption{The two panels show the integrands of $C_{\rm h}$ and
    $C_{\rm v}$, respectively, and the areas colored in red represent
    $C_{\rm h}$ and $C_{\rm v}$ themselves. Both integrals have a
    finite point integrable singularity (the exponents $\beta_{\rm h}$
    and $\beta_{v}$ are in the interval $(0,1)$) at their left
    boundaries: the insets show the asymptotic power-law scaling of
    the singularities.
\label{fig:15}}
\end{figure}
Then, choosing a value $\mu_{\rm IR}=1/16$ to delimit the infrared region for vertical wave numbers, we
have
\begin{equation}\label{eq:5.17}
\begin{aligned}
	C_{\rm v} \simeq \int_1^{\frac{m_{\max}}{m_{\min}}} \mu^{-2}
        \mathcal{K_{\rm diss}(\mu)} &\simeq -\frac{d_{\rm v}
          \,{\mu_{\rm IR}}^{1-\beta_{\rm v}}}{1-\beta_{\rm v}} +
        \int_{1+\mu_{\rm IR}}^{\frac{m_{\max}}{m_{\min}}} d\mu \,\mu^{-2}
        \mathcal{K}_{\rm diss}(\mu)\\ &\simeq - 8\pi(18.0 + 14.0 + 3.1) \simeq
        -8\pi\times 35.1\,.
\end{aligned}
\end{equation}
Again, the numbers $18.0$, $14.0$ and $3.1$ represents infrared (ID), near-colinear and unclassified contributions, respectively, with a negligible contribution from the ultraviolet regions.

Now, we compute the power crossing the $k_{\max}$ boundary,
denoted as $P_{\rm h}$ (horizontal boundary), and the power crossing
the $m_{\max}$ boundary, denoted as $P_{\rm v}$ (vertical boundary), via
\begin{equation}\label{eq:pow}
\begin{aligned}
&P_{\rm h} = \int_{m_{\rm min}}^{m_{\rm max}} F_{\rm diss}(m) \;dm =
  D_{\rm h}(1-s)^{-1} \left(\frac{N}{\alpha}\right)^{-s} (m_{\rm max}^{1-s} - m_{\rm
    min}^{1-s}) \,,\\ &P_{\rm v} =
  \int_{\frac{f}{\alpha}m_{\max}}^{\frac{N}{\alpha}m_{\max}} G_{\rm
    diss}(k) \;dk = D_{\rm v} s^{-1}
  \left[\left(\frac{f}{\alpha}\right)^{-s}-\left(\frac{N}{\alpha}\right)^{-s}\right]m_{\rm
      max}^{1-s} \,,
\end{aligned}
\end{equation}
where
\begin{equation}\label{eq:fluxnonloc}
	D_{\rm h} = {\color{black}4}\pi \frac{{{\color{black}N}}^2}{g}(V_0 A)^2 C_{\rm h}\,\quad D_{\rm v} = {\color{black}4}\pi \frac{{{\color{black}N}}^2}{g}(V_0 A)^2 C_{\rm v}\,,
\end{equation}
Using \eqref{eq:64}, \eqref{eq:A}, \eqref{eq:pow} and \eqref{eq:fluxnonloc}, we obtain
\begin{equation}\label{eq:pownonloc}
\begin{aligned}
&P_{\rm h} = \frac{\Gamma C_{\rm h}}{1-s}\;{\color{black} \left[1-\left(\frac{\ell}{2b}\right)^s\right]}\;{\color{black}f^{1+s}} N^{2}
  E^2\,,\\ &P_{\rm v} = \frac{\Gamma C_{\rm v}}{s}\;
 \left[1-\left(\frac{f}{N}\right)^s\right]\; f {\color{black}N^{1+s}} E^2\,, \\ &
	\Gamma = {\color{black}{\frac{3}{4\pi^3}}\left(\frac{{3}\ell}{2b}\right)^s
        \frac{\rho_0 b^2{N_0^{1-s}}}{g\ell}\,, \qquad
        s=0.38\,, \quad \ell=10{\rm\, m} }.&
\end{aligned}
\end{equation}
%
%

%
\begin{figure}
  \centerline{
    \includegraphics[width=\linewidth]{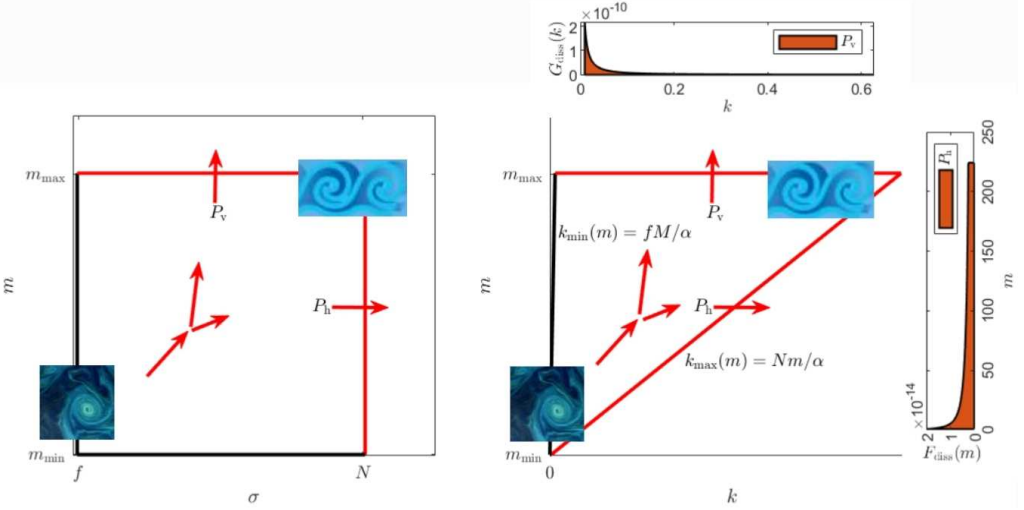}
  }
  \caption{Left: the {\it inertial box} in $\sigma-m$ space. Right: the {\it inertial box} in $k-m$ space. The low frequency energy forcing is represented in the picture a mesoscale eddy, while wave breaking (with generation of turbulent vortices) represents dissipation at high wave numbers. The distribution of energy dissipation along the absorbing boundaries is depicted in the two insets showing $F_{\rm diss}(m)$ and $G_{\rm diss}(k)$, whose integrals (red area) give respectively $P_{\rm h}$ and $P_{\rm v}$, Eq. \eqref{eq:pow}. In these two plots, $k$ is in units of ${\rm m}^{-1}$, $m$ is in units of ${\rm m}^3{\rm kg}^{-1}$, $G_{\rm diss}(k)$ is in $\rm {W}\,{\rm m} \, {\rm kg}^{-1}$, $F_{\rm diss}(m)$ is in $\rm {W}\, {\rm m}^{-3}$, so that both integrals give a power per unit mass.
\label{fig:16}}
\end{figure}
In Fig. \ref{fig:16} a graphical interpretation of the computations in
the above paragraph is shown. The physical cutoffs define a
rectangular box (left panel) within which energy is transfered through
resonant interactions toward high vertical wave numbers and high
horizontal wave numbers. If energy is provided by large scale forcing
at low frequency, for simplicity represented as a mesoscale eddy in
the figure, it ends up being dissipated at high frequencies or high
wave numbers, represented as wave breaking with generation of
turbulent vortices. In $k-m$ space this {\it inertial box} translates
into a trapezoid whose lateral sides are straight lines defined by the
dispersion relation. The integrals in Eq. \eqref{eq:pow} quantify the
contribution to the dissipated power along the dissipative sides of
the box. In the figure, the red areas quantify how dissipation is
distributed along these dissipative sides of the inertial box.

Eq. \eqref{eq:pownonloc}   is the main result of our paper. The outgoing
power toward large wave numbers is the quantity that is modelled by
the finescale parametrization formula of ocean
mixing~[\cite{polzin1995finescale}] as
\begin{equation}\label{eq:fs0}
	P_{\rm fs} = (1.9 \times 10^7 \,\text{\rm m}^2)\, \cosh^{-1}\left(\frac{N}{f}\right)\, f N^2 E^2\,,
\end{equation}
which is in agreement with \eqref{eq:pownonloc} regarding the
dependence upon the main physical parameters, if the lower-order
corrections $(f/N)^{s}$ and $\cosh^{-1}(N/f)$ are neglected.  The
predicted intensities can also be compared directly, using the
standard parameters of the GM spectrum \eqref{eq:GM} with $N=N_0$,
which yields
\begin{equation}\label{eq:pownonlocnum}
\begin{aligned}
&P_{\rm h} \simeq - {\color{black}0.5} \times 10^{-8}\, \text{\rm W  kg}^{-1}\,,\\
&P_{\rm v} \simeq -{\color{black}1.5} \times 10^{-8}\, \text{\rm W  kg}^{-1}\,
\end{aligned}
\end{equation}
This amounts to a total dissipated power
\begin{equation}
	P_{\rm tot} = P_{\rm h} + P_{\rm v} \simeq - {\color{black}2.1} \times 10^{-8}\, \text{\rm W kg}^{-1}\,,
\end{equation}
which is negative since it is lost by the wave system considered.
With the same parameters and the same sign convention, the finescale
parametrization formula predicts a total dissipated power
\begin{equation}\label{eq:fs}
	P_{\rm fs} \simeq - 0.8 \times 10^{-9}\, \text{\rm W kg}^{-1}\,.
\end{equation}
These two predictions are in qualitative agreement, but with a difference
of about an order of magnitude.
\footnote{Had we used a
  factor of $4$ in place of $3$ in the definition of $m_\star$,
  Eq.~\eqref{eq:GM}, the fluxes would have to be multiplied by a
  factor of $(4/3)^{1+s}\simeq1.5$, which does not affect the order of
  magnitude of the estimate. Here, we
      use the original choice of $m_\star$ from [\cite{GM76}] for simplicity. We hope to consider oceanographic
      implications of this choice in details in future publications.}}
  We elucidate on possible reasons in the conclusions. Moreover, the
  following remarks are important for the sake of clarity.

{Above, we have used the terminology {\it dissipated power} to indicate the amount of energy that is absorbed by the boundary sink at small scales. As far as the internal wave kinetic equation is concerned, this sink is acting as an actual dissipation term. In the finescale parametrization picture to which we refer (see~[\cite{polzin2014finescale}], Sec. 2), this amount of energy is converted into turbulent kinetic energy at small scales, and therefore roughly equals the turbulent energy production. In turn, the latter contributes separately both to diapycnal mixing and to heat production, in proportions that can be quantified by closure assumptions beyond the scope of the present paper. We stress that the use of the word {\it dissipation} in this paper does not refer to dissipation of energy as heat, but to the role played by the sink that takes energy out of the wave system considered.}

{The energy fluxes at the boundaries, Eqs.~\eqref{eq:Fdiss}-\eqref{eq:Gdiss}, show a non-trivial dependence on the variables $k$ and $m$. In non-isotropic systems, rather than one single stationary state it is common to have a family of stationary states ([\cite{NazBook}], Sec. 9.2.5) whose energy fluxes depend on $k$ and $m$ in different ways. The generalized Kolmogorov-Zakharov spectrum (that is the Pelinovski-Raevski spectrum, for internal waves) is only one among the members of the family. Then, the locality conditions imply that for the internal waves only one solution is physically meaningful, i.e. $(a,b)=(3.69,0)$. Indeed, since the solution is stationary, for any enclosed region in $k-m$ space
(contained in the inertial range where no sources or sinks are
present), the incoming energy flux must be equal to the outgoing energy flux (and opposite in sign). Equivalently, as long as both ends of a boundary are fixed, the total outgoing flux through the boundary must be independent of the path of the boundary in $k-m$ space. For instance, one could show that the flux through the dissipative boundary equals the flux through the forcing boundary (respectively, the red and the black boundaries in Fig.~\ref{fig:16}).}

{Finally, we acknowledge the fact that the solutions with
  $b=0$ are known in the literature as {\it no flux} solutions to the
  Fokker-Planck (diffusive) approximation to the kinetic
  equation~\eqref{eq:11}. How this approximation is achieved is shown
  for instance in~[\cite{iwthLPTN}], by use of a straightforward
  leading order approximation of the infrared Induced Diffusion
  interactions. An exact leading-order balance makes the infrared flux
  apparently vanish, whose trace in this paper is found in the two
  exact cancellations in the infrared region. However, our
  computations leading to Eq.~\eqref{eq:30} show how the next leading
  order terms carry a small but non-zero flux toward high horizontal
  wave numbers. Moreover, a net energy flux is indeed due to
  interactions outside the infrared region, such as the colinear
  interactions. Therefore, there is no mathematical contraddiction in having $b=0$
    and a nonzero downscale energy flux. Indeed, the considerations in
    the present paper go way beyond the {\it diffusive
      approximation} to the internal wave kinetic equation, where the
    concept of {\it no flux solutions} was concocted.}
\section{Discussion and conclusions}\label{sec:conclusions}
This paper is focused on the specific case of a scale invariant field
of internal waves in the ocean with vertically homogeneous ($b=0$) wave
action. For such a case we have found that

$\bullet$
 There necessarily exists a stationary state for  $3 < a < 4$ dictated
  by the opposite signs of the infrared and ultraviolet resonant singularities.

  $\bullet$
The dominant contributions to the integrand of the kinetic equation are coming from
  Induced Diffusion  and near-colinear resonances;

  $\bullet$
 Both near-colinear and extreme-scale-separated resonances are
  subject to subtractive cancellation between differing triad types;

  $\bullet$
 Both types of resonances balance to zero for $a \simeq 3.69$
  independently, cf. Fig. \ref{fig:11a}{\bf (A)};

  $\bullet$
 The contribution from the unclassified resonances is negligible,
  although it balances to zero at $a \simeq 3.69$, too;

  $\bullet$ We point out an important role played by colinear
  resonances, and we find a {\it modified Pelinovsky-Raevsky} spectrum,
  Eq. (\ref{ThreePointFive}), from a balance of the largest colinear contributions. Curiously, this solution is precisely in
  {the middle of the interval of values of the exponent for which the collision integral is convergent}, [\cite{ZLF}]. 
 With the help of this analytical result, we explained how the exponent $3.69$ appears concerning the near-colinear region. It is
  a result of a shift of the balance of the {\it modified Pelinovsky-Raevsky} spectrum due to the negative contribution from the subleading colinear triads.

$\bullet$ By considering the sign of the integrand of the kinetic equation we
  can visualize a downscale flux of energy in the horizontal wave number space, see Fig. \ref{fig:6}.

  $\bullet$ Modifying the scale invariant spectrum (\ref{eq:3.69}) to include
  natural physical boundaries in Fourier space we numerically calculate the
  value of the spectral flux of energy towards high horizontal and vertical wave numbers.
  This {\it quantitative} estimate might be useful for a direct estimate of turbulent energy mixing in the ocean.

  $\bullet$ Our results compares favorably with the {\it finescale
    parametrization} formula of [\cite{polzin1995finescale}],
  reproducing the correct order of magnitude.

  Our results are obtained for a scale invariant internal wave field
  with vertically homogeneous wave action profile. This profile is
  close to the scale invariant limit of the Garrett and Munk spectrum,
  yet it is a strong idealization of the actual internal wave
  spectrum. Generalizing the evaluations presented in this paper for
  more realistic ocean internal wave spectra, {including the presence of background rotation and spatial inhomogeneity}, is a subject of current
  research.

  \acknowledgments {{\bf Supplementary data.}} Supplementary material is found in attachment to the present file.

  \acknowledgments {{\bf Acknowledgments}} The authors are grateful to
  Dr. Kurt Polzin for multiple discussions. Support from Office of
  Naval Research grant N00014-17-1-2852 is gratefully
  acknowledged. YL gratefully acknowledges support from NSF
      DMS award 2009418.

  \acknowledgments {{\bf Declaration of Interests.}} The authors report no conflict of interest.

\bibliographystyle{jfm}
\bibliography{ref}
\newpage
\pagebreak
\setcounter{page}{1}

\begin{center}
{\huge Supplementary material}
\end{center}
\section{Infrared and ultraviolet singularities}\label{app:AA}

\subsection{Infrared singularity}

\begin{figure}
\centerline{\includegraphics[width=0.6\linewidth]{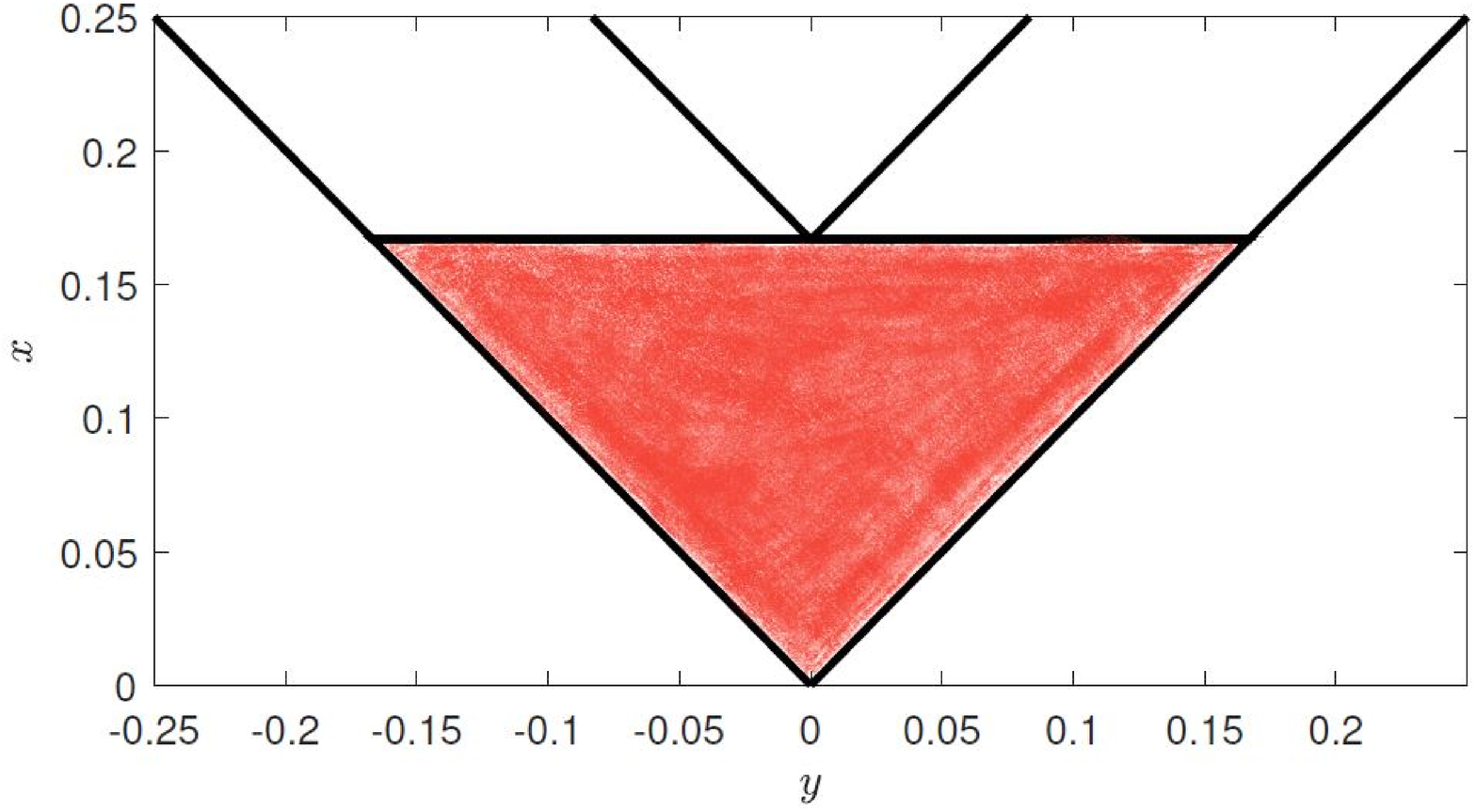}}
\centerline{\includegraphics[width=0.5\linewidth]{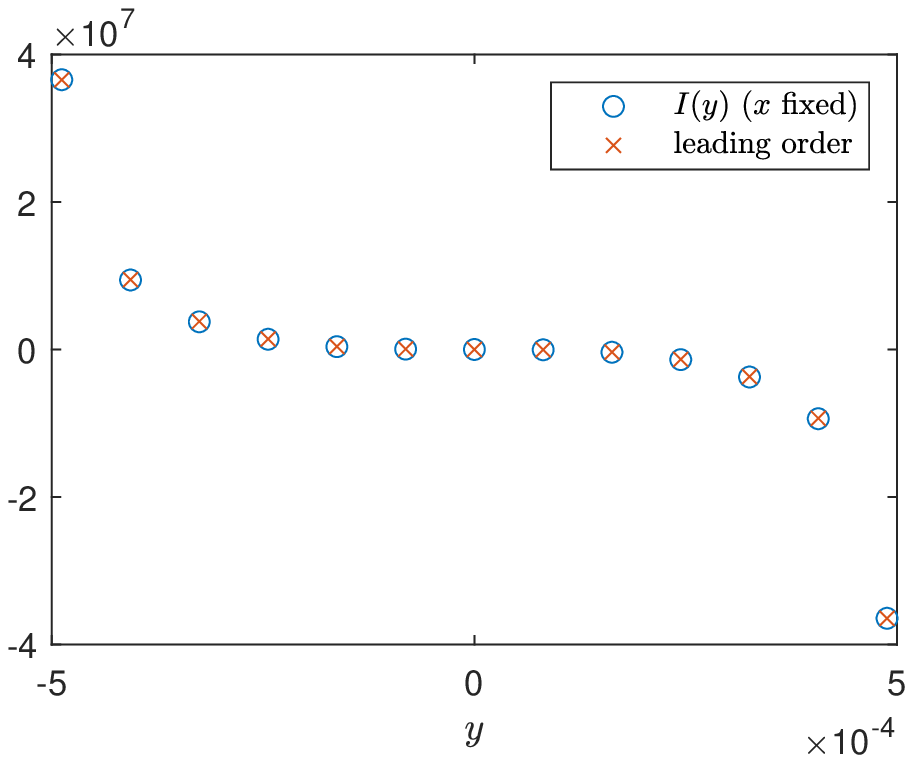}%
\includegraphics[width=0.5\linewidth]{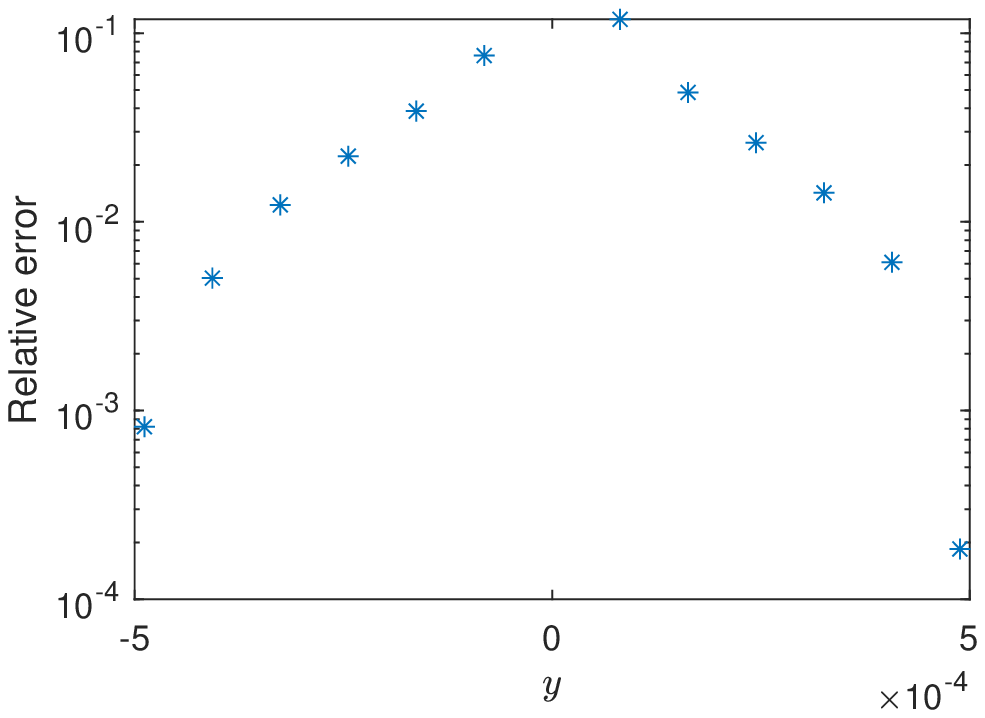}}
  \caption{Top: Sub-region ${\rm IR}$, containing the singularity $k_2\to0$. The region is represented after performing the change of variables $(k_1,k_2)\to(y,x)$.
Bottom left: Some values of the integrand of~\eqref{eq:29} during the double loop of integration in the $x$ and $y$ variables. The circles are the full integrand integrated numerically, and the crosses are the leading order of the integrand in~\eqref{eq:29}. In the plot, $x$ is fixed and $y$ varies from $-x$ to $x$. Note the magnitude of the integrand as the singular boundary is approached.  Bottom right: relative error of the points in the left panel, in logaritmic scale.
 \label{fig:2}}
\end{figure}

\noindent In the region ${\rm IR}$, in red in Fig.~\ref{fig:A1}, let us change variables using
\begin{equation}
	k_1=k(1+y),\qquad k_2= kx\,,
\end{equation}
which gives the new integration area shown in the left panel of Fig.~\ref{fig:2}, with $x,y=O(\epsilon), \epsilon\ll1$.
Now, using a power-series expansion the factor $f^0_{12}$, we notice that the dominant matrix elements are those given by the resonances (${\rm Ia}$) and (${\rm IIa}$), the Induced Diffusion contributions, and we obtain
\begin{equation}\label{eq:22}
	\begin{aligned}
		R^0_{12} f^0_{12} -& R^1_{02} f^1_{02} - R^2_{01} f^2_{01} \\
		\simeq &\; R^k_{k(1+y),kx}[n_{k(1+y)}n_{kx} - n_k(n_{k(1+y)}+n_{kx})] - R^{k(1+y)}_{k,kx}[n_{k}n_{kx} - n_{k(1+y)}(n_k+n_{kx})] \\
		\simeq &\; n_{kx} \frac{\partial n_k}{\partial k} ky \left[R^k_{k(1+y),kx} + R^{k(1+y)}_{k,kx}\right]\\
		\simeq &\; -  a k^{-2a} x^{-a}y  \left[R^k_{k(1+y),kx} + R^{k(1+y)}_{k,kx}\right]\,,
	\end{aligned}
\end{equation}
where $R^2_{01}$ has been neglected, and the so-called {\it first cancellation} has taken place due to the fact that $\lim_{k\to0}n_{k(1+y)} = n_k$. To evaluate the terms $R^k_{k(1+y),kx}$ and $ R^{k(1+y)}_{k,kx}$ we notice that the two leading-order contributions are given by the conditions $({\rm Ia})$ and $({\rm IIa})$ where, respectively,
\begin{equation}
	m_1^\star \simeq m(1+\sqrt{x}),\qquad m_2^\star \simeq -m(\sqrt(x)+\frac12 (x+y))\,,
\end{equation}
\begin{equation}
	m_1^\star \simeq m(1-\sqrt{x}),\qquad m_2^\star \simeq -m(\sqrt(x)-\frac12 (x+y))\,.
\end{equation}
Thus, we obtain
\begin{equation}\label{eq:25}
	|V^k_{k(1+y),kx}|^2 \simeq 4k^3m^{-1} \left[ \frac{y^2}{x^2}\sqrt{x} + \frac{y^2}{2x^2}(x+y) - y \right]\,,
\end{equation}
\begin{equation}\label{eq:26}
	|V^{k(1+y)}_{k,kx}|^2 \simeq 4k^3m^{-1} \left[ \frac{y^2}{x^2}\sqrt{x} + \frac{y^2}{2x^2}(x+y) + y \right]\,.
\end{equation}
Furthermore, we have
\begin{equation}\label{eq:27}
	|{g^0_{12}}'|\simeq |{g^1_{02}}'|  \simeq m^2/{2k}\,,\qquad\Delta_{012} \simeq \frac{k^2}{\sqrt{x^2-y^2}}\,.
\end{equation}
Using Eqs.~\eqref{eq:25}-\eqref{eq:27}, we finally obtain
\begin{equation}
	R^k_{k(1+y),kx} + R^{k(1+y)}_{k,kx} \simeq 2k^3mx \left[2\frac{y^2}{x^2}\sqrt{x} + \frac{y}{x^2}(y^2-x^2)\right]/\sqrt{x^2-y^2}\,.
\end{equation}
Thus, Eq.~\eqref{eq:22} yields a contribution $\mI_C(k)$ to the total $\mI(k)$ (due to integration in region C):
\begin{equation}\label{eq:29}
\begin{aligned}
	\mI_{\rm IR}(k) &= 8\pi k\int_{\rm IR} dx\, dy\;  \left[ -  a k^{-2a} x^{-a}y  \left(R^k_{k(1+y),kx} + R^{k(1+y)}_{k,kx}\right)\right]\\
			& = -16\pi a k^{-2a+4} m \int_0^{x_{\rm IR}} dx \int_{-x}^x dy\; x^{-a-1} \frac{2y^3\sqrt{x} + y^2(y^2 -x^2)}{\sqrt{x^2 - y^2}}\,.
\end{aligned}
\end{equation}
The first term of the integrand is dominating but it is an odd function of $y$ in a domain symmetric with respect to the origin, so its contribution is vanishing: this is the {\it second cancellation}, implying:
\begin{equation}\label{eq:30A}
\begin{aligned}
	I_{\rm IR}(k) & \sim -16\pi a k^{-2a+4} m \int_0^{x_{\rm IR}} dx \int_{-x}^x dy\; x^{-a-1} \frac{y^2(y^2-x^2)}{\sqrt{x^2 - y^2}}\\
			& = 2\pi^2 \frac{a}{4-a}m k^{-a }{k_{\rm IR}}^{-a+4}\,,
\end{aligned}
\end{equation}
where $k_{\rm IR}$ is the (small) height of the red region in Fig.~\ref{fig:2}.
For convergence of the integral we have to impose that $a<4$, and we notice that the contribution is positive.

The leading order of the integrand of Eq.~\eqref{eq:29} is tested numerically in the right panel of Fig.~\ref{fig:2}, leaving no doubt about its correctness.
\subsection{Ultraviolet singularity}
\noindent %
\begin{figure}
\centerline{\includegraphics[width=0.45\linewidth]{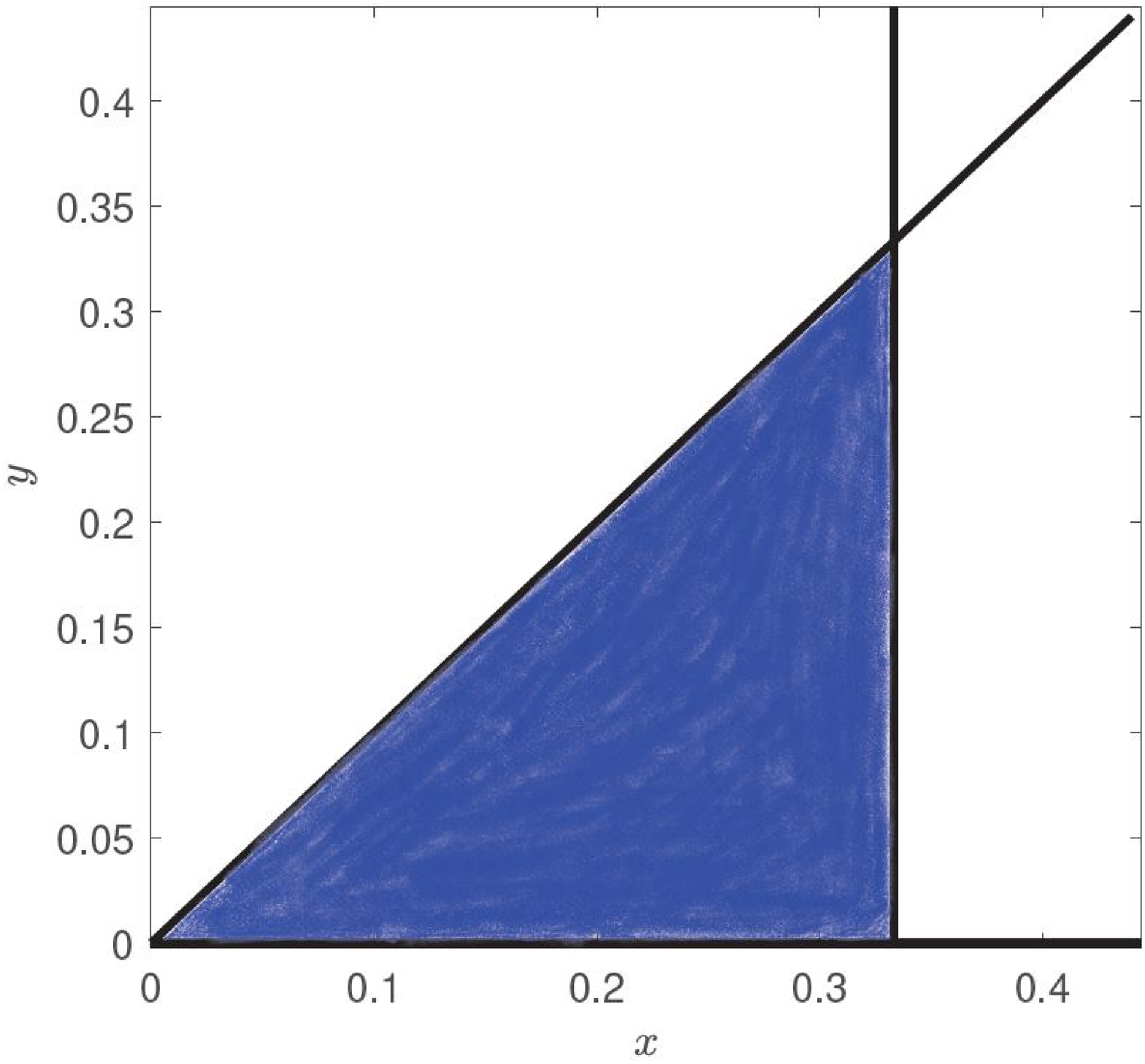}}
\centerline{\includegraphics[width=0.5\linewidth]{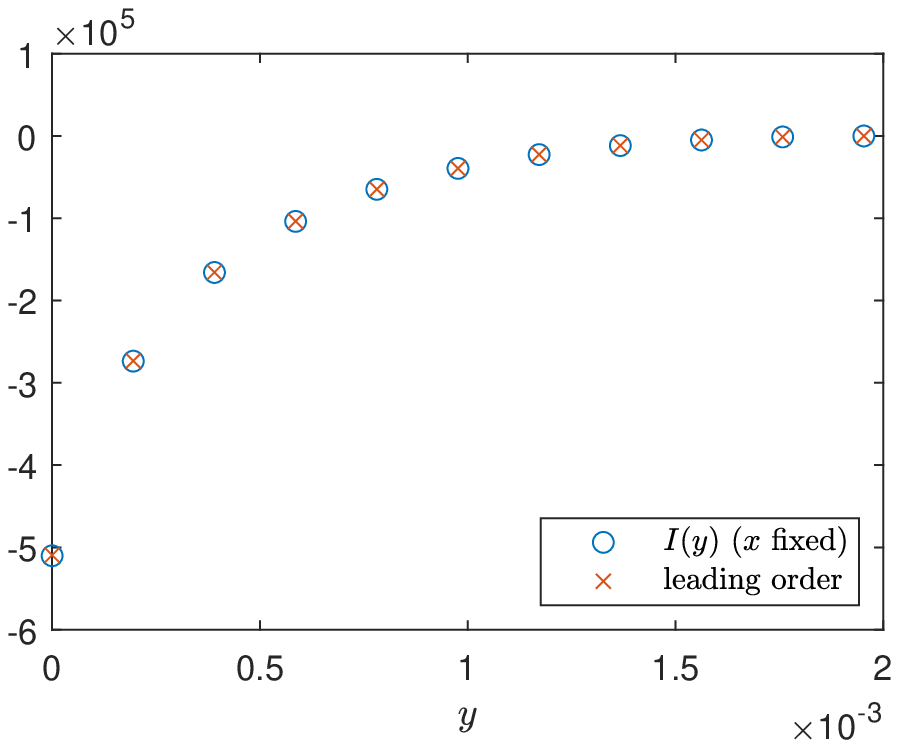}%
\includegraphics[width=0.5\linewidth]{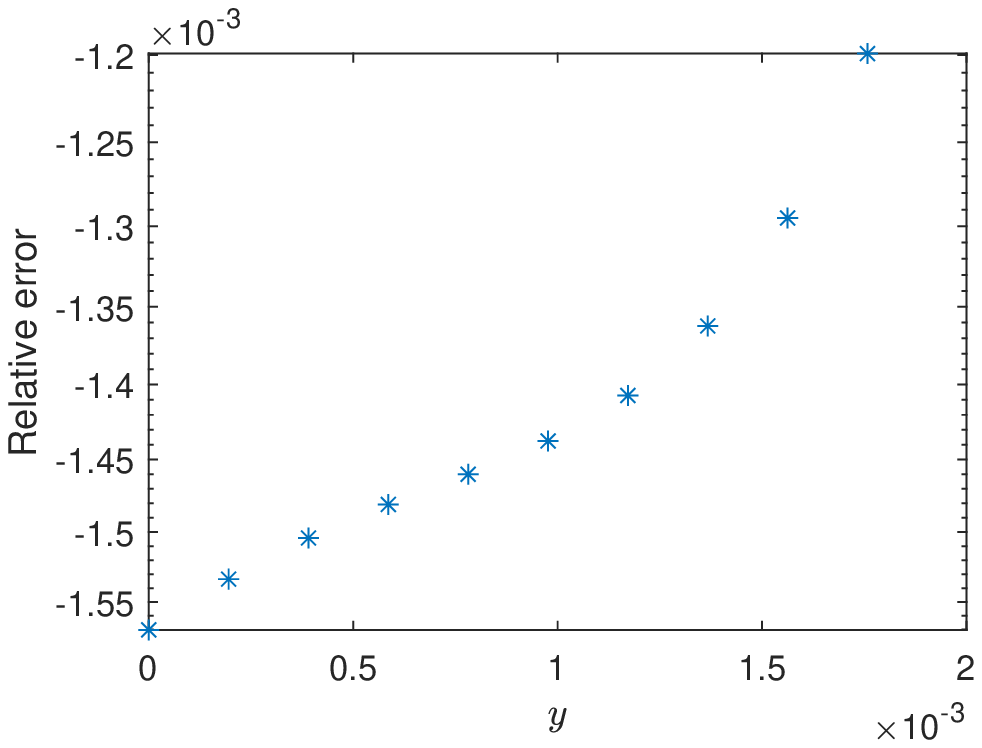}}
  \caption{Top: Sub-region $D$, in the variables defined by Eq.~\eqref{eq:31}. 
Bottom left: Some values of the integrand of~\eqref{eq:38} during the double loop of integration in the $x$ and $y$ variables. In the plot, $x$ is fixed and $y$ varies from $-x$ to $x$. Note the magnitude of the integrand as the singular boundary is approached.  Bottom right: relative error of the points in the left panel, in logaritmic scale.
 \label{fig:3}}
\end{figure}
Let us focus on the region ${\rm UV}$, applying the following change of variables:
\begin{equation}\label{eq:31}
	k_1 = \frac{k}{x},\qquad k_2 = \frac{k}{x}(1+y-x)\,,
\end{equation}
for $x\ll1$  and $y\in(0,x)$\,.
$1/x$ provides the largeness of $k_1$ and $k_2$, while $y=0$ corresponds to the singular line $k_2=k_1-k$, thereby allowing singular power expansions around the point $y=0$. Finally, $y=x$ corresponds to the line $k_2=k_1$. In the new variables, the region ${\rm UV}$ is represented in Fig.~\ref{fig:3} (for a value $k_{\rm UV}=3k$).
We have that the integrand of the {\rm UV} contribution is given by
\begin{equation}\label{eq:32}
	\begin{aligned}
		R^0_{12} f^0_{12} -& R^1_{02} f^1_{02} - R^2_{01} f^2_{01} \\
		\simeq &\; -R^{k/x}_{k,k(1+y-x)/x}[n_{k}n_{k(1+y-x)/x} - n_{k/x}(n_{k}+n_{k(1+y-x)/x})] \\
					& - R^{k(1+y-x)/x}_{k,k/x}[n_{k}n_{k/x} - n_{k(1+y-x)/x}(n_k+n_{k/x})] \\
		\simeq &\; -a (x-y) k^{-2a}x^a \left[R^{k/x}_{k,k(1+y-x)/x} - R^{k(1+y-x)/x}_{k,k/x}\right]\,,
	\end{aligned}
\end{equation}
where $R^0_{12}$ has been neglected (subleading), and the {\it first cancellation} has taken place in the terms $f^1_{02}$ and $f^2_{01}$, expanded to first order in $x$ and $y$.
Now, the two leading order contributions are given by the matrix elements satisfying the resonance conditions (${\rm IIb}$) and (${\rm IIIb}$), again the Induced Diffusion contributions, yielding respectively
\begin{equation}
	m_1^\star \simeq -m \left( \frac{1}{\sqrt{x}} - \frac{y}{2x} \right)\,,\qquad m_2^\star \simeq -m \left( \frac{1}{\sqrt{x}} + 1 - \frac{y}{2x} \right)\,,
\end{equation}
\begin{equation}
	m_1^\star \simeq -m \left( \frac{1}{\sqrt{x}} + \frac{y}{2x} \right)\,,\qquad m_2^\star \simeq -m \left( \frac{1}{\sqrt{x}} - 1 + \frac{y}{2x} \right)\,,
\end{equation}
which imply the leading order expressions
\begin{equation}
	|V^1_{02}|^2/|{g^2_{13}}'| \simeq \frac{2k^2m (x-y)^2}{x^{9/2}} + \frac{2k^2m(2x^3-4x^2y+3xy^2-y^3)}{x^5}\,,
\end{equation}
\begin{equation}
	|V^2_{01}|^2/|{g^2_{01}}'| \simeq \frac{2k^2m (x-y)^2}{x^{9/2}} - \frac{2k^2m(2x^3-4x^2y+3xy^2-y^3)}{x^5}\,.
\end{equation}
Plugging into the expression~\eqref{eq:22}, with
\begin{equation}
	\Delta_{012}\simeq \frac{k^2}{x^2}\sqrt{(2x-y)y}\,,
\end{equation}
we finally obtain the following expression for the integrand:
\begin{equation}\begin{aligned}\label{eq:38}
	 -a (x-y) k^{-2a}x^a& \left[R^{k/x}_{k,k(1+y-x)/x} - R^{k(1+y-x)/x}_{k,k/x}\right] \\
&=  -a (x-y) k^{-2a}x^a \frac{k^3}{x^2} \left[\frac{4k^2m}{x^5} (2x^3-4x^2y+3xy^2-y^3)\right] \frac{x^2}{k^2\sqrt{(2x-y)y}}\\
&= -4a k^{-2a+5} m x^{a-8} \left[ (x-y)^4 + x^2(x-y)^2 \right]/\sqrt{(2x-y)y}\,.
\end{aligned}\end{equation}
This can be integrated analytically, giving the following result:
\begin{equation}\label{eq:39A}
\begin{aligned}
	\mI_{\rm UV}(k) &= -32\pi a k^{-2a + 4}m\int_0^{x_{\rm UV}} dx \int_0^x dy\; \frac{k^2}{x^3} \;x^{a-8} \left[ (x-y)^4 + x^2(x-y)^2 \right]/\sqrt{(2x-y)y}\\
&\simeq -14 \pi^2 \frac{a}{a-3} k^{-a+1} m\; k_{\rm UV}^{3-a}\,.
\end{aligned}
\end{equation}
For convergence of the integral we must have that $a>3$, which along with the infrared condition results into the convergence segment $3<a<4$. Evidence of the correctness of the convergence condition $-3<a<4$ is given in the figures in the next Section, where the convergent integral is computed numerically.

\section{Numerical computation of integrable singularities}\label{appA}
\begin{figure}
\centerline{\includegraphics[width=0.5\linewidth]{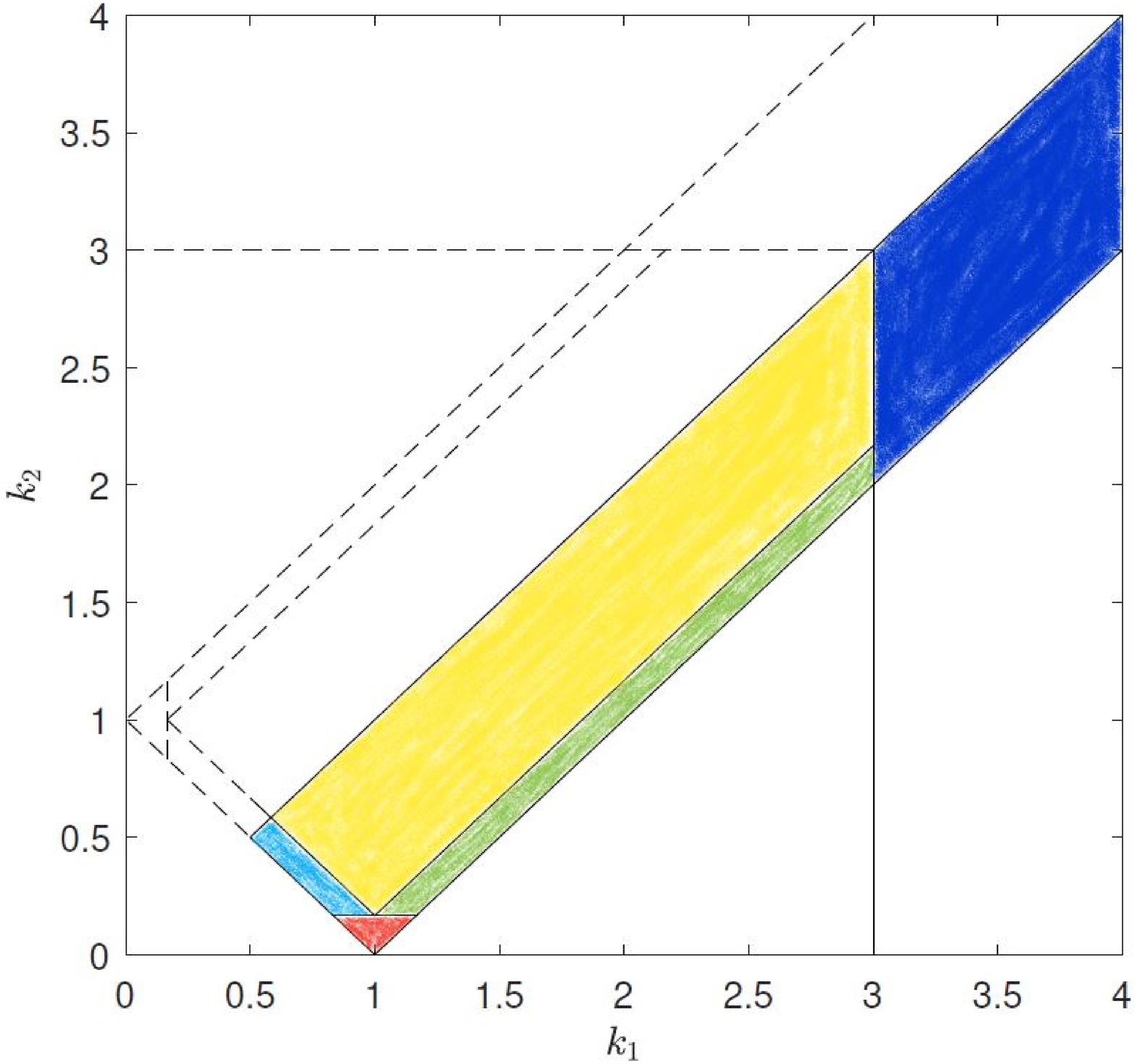}}
  \caption{The five subregions of the kinematic box, where integration is performed (in the figure, $k=1$): region $\mR_1$ (or ${\rm IR}$ in the main text) in red, region $\mR_2$ in light blue, region $\mR_3$ in green, region $\mR_4$ in yellow, and region $\mR_5$ (or ${\rm UV}$ in the main text) in blue. $\mR_1$ contains the singularity for $k_2\to0$; $\mR_2$ and $\mR_3$ contain the singularities due to a vanishing {\it area of the interaction triangle with sides $k$, $k_1$ and $k_2$}, which is at denominator in the integrand; $\mR_4$ contains no singularities of the integrand; $\mR_5$ is integrated over after a suitable change of coordinates that map the singularity of the integrand approaching infinity into the origin in the new coordinate system.
 \label{fig:A1}}
\end{figure}
The five regions represented in Fig.~\ref{fig:A1}, identified by the values of the two parameters $k_{\rm IR}$ and $k_{\rm UV}$, are treated separately. Just for numerical purpose, it is easier to consider these new regions, and then stich the results together to reconstruct the contributions of each of the regions in the main text.  We recall that $k_{\rm IR}$ defines the height of the red region in Fig.~\ref{fig:2} in the $k_2$ direction, and $k_{\rm UV}$ defines the lower boundary of the blue region in the $k_1$ direction. The singularities are integrated with the following procedure: the divergent leading order is subtracted to the integrand, giving a convergent difference that is integrated numerically. Subsequently, the exact analytical result of the integral of the subtracted part is added, ensuring convergence. The leading order in the regions $\mathcal R_1$ and $\mathcal R_5$ is given in the previous section, while the regions $\mathcal R_2$ and $\mathcal R_3$ are singular because of the $\Delta_{012}$ denominator at the boundary, i.e. the vanishing area of triangles formed by triads of collinear horizontal wave vectors.

In the code, the five terms are decomposed and integrated in the following form:
\begin{itemize}
\item Region $\mR_1$:
\begin{equation}
	\mI_1 = \mL + k^2 \int_0^{k_{\rm IR}/k} dx' \int_{-x}^x dy \left[ J(x,y) - L(x,y) \right]\,,
\end{equation}
where
\begin{equation*}
	J(x,y) = R^k_{k(1+y),kx} f^k_{k(1+y),kx} - R^{k(1+y)}_{k,kx}f^{k(1+y)}_{k,kx} - R^{kx}_{k,k(1+y)}f^{kx}_{k,k(1+y)}\,,
\end{equation*}
\begin{equation*}
	L(x,y) = -2a k^{-2a+3} m x^{-a-1} \frac{2y^3\sqrt{x} + y^2(y^2 -x^2)}{\sqrt{x^2 - y^2}}\,,\qquad \mL = \frac\pi4 \frac{a}{4-a}m k^{-a +1}k_{\rm IR}^{-a+4}\,.
\end{equation*}
\item Region $\mR_2$:
\begin{equation}\label{eq:41}
	\mI_2 = \int_{k/2}^{k-k_{\rm IR}} dk_1 \left[ \mL(k_1) +  \int_0^{k_{\rm IR}} dx \left( J(x,k_1) - L(x,k_1) \right) \right] \,,
\end{equation}
where
\begin{equation*}
	J(x,k_1) =  \frac{T^k_{k_1, k-k1+x} - T^{k_1}_{k,k-k_1+x} - T^{k-k_1+x}_{k,k_1}}{\Delta_{k,k_1,k-k_1+x}}\,,
\end{equation*}
\begin{equation*}
	L(x,k_1) =  \frac{T^k_{k_1, k-k_1} - T^{k_1}_{k,k-k_1} - T^{k-k_1}_{k,k_1}}{\sqrt{2kk_1(k-k_1)x}}\,, \qquad \mL(k_1) =  \left( T^k_{k_1, k-k_1} - T^{k_1}_{k,k-k_1} - T^{k-k_1}_{k,k_1} \right) \frac{\sqrt{2k_{\rm IR}}}{\sqrt{k k_1(k-k_1)}}\,,
\end{equation*}
and the following definition was used:
\begin{equation*}
	T^0_{12} = kk_1k_2 |V^0_{12}|^2 f^0_{12}/|{g^0_{12}}'|\,.
\end{equation*}
\item Region $\mR_3$:
\begin{equation}\label{eq:42}
	\mI_3 = \int_{k+k_{\rm IR}}^{k_{\rm UV}} dk_1 \left[ \mL(k_1) +  \int_0^{k_{\rm IR}} dx \left( J(x,k_1) - L(x,k_1) \right) \right] \,,
\end{equation}
where
\begin{equation*}
	J(x,k_1) =  \frac{T^k_{k_1, k1-k+x} - T^{k_1}_{k,k_1-k+x} - T^{k_1-k+x}_{k,k_1}}{\Delta_{k,k_1,k_1-k+x}}\,,
\end{equation*}
\begin{equation*}
	L(x,k_1) =  \frac{T^k_{k_1, k_1-k} - T^{k_1}_{k,k_1-k} - T^{k_1-k}_{k,k_1}}{\sqrt{2kk_1(k_1-k)x}}\,, \qquad \mL(k_1) =  \left( T^k_{k_1, k_1-k} - T^{k_1}_{k,k_1-k} - T^{k_1-k}_{k,k_1} \right) \frac{\sqrt{2k_{\rm IR}}}{\sqrt{k k_1(k_1-k)}}\,,
\end{equation*}
\item In region $\mR_4$ the integrand is finite and integration is straightforward:
\begin{equation}
	\mI_4 = \int_{\mR_4} dk_{12} \; (R^0_{12}f^0_{12} - R^1_{02}f^1_{02} - R^2_{01}f^2_{01})\,.
\end{equation}
\item Region $\mR_5$:
\begin{equation}\label{eq:44}
\mI_5 =\mL + \int_0^{k/k_{\rm UV}} dx \int_0^x dy \left[ J(x,y) - L(x,y) \right]\,,
\end{equation}
where
\begin{equation*}
	J(x,y) = \left[R^k_{k/x,k(1+y-x)/x} f^k_{k/x,k(1+y-x)/x} - R^{k/x}_{k,k(1+y-x)/x}f^{k/x}_{k,k(1+y-x)/x} - R^{k(1+y-x)/x}_{k,k/x}f^{k(1+y-x)/x}_{k,k/x} \right] \frac{k^2}{x^3}\,,
\end{equation*}
\begin{equation*}
	L(x,y) = -4a k^{-2a+5} m x^{a-8} \left[ (x-y)^4 + x^2(x-y)^2 \right]/\sqrt{(2x-y)y}\,,\qquad \mL = -\frac74 \pi \frac{a}{a-3} k^{-a+2} m\; k_{\rm UV}^{3-a}\,.
\end{equation*}
\end{itemize}

\section{Numerical convergence and independence from the cuts}\label{appB}
\begin{figure}
  \includegraphics[width=\linewidth]{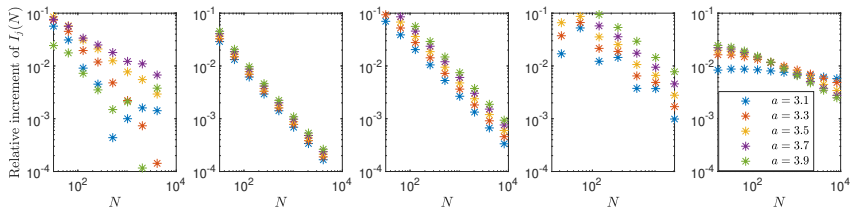}
  \caption{Test of convergence in each subregion of integration, as the number of gridpoints ($N$) increases. From left to right, the plots refer to regions from $\mathcal R_1$ to $\mathcal R_5$, respectively, and the plotted quantity is the absolute value of the relative increment of each contribution, as $N$ doubles (in log-log scale). All contributions appear to converge up to variations of at most $1\%$, for the largest numbers of points here considered.
 \label{fig:4}}
\end{figure}

In Fig.~\ref{fig:4} such numerical convergence is shown independently for each of the five regions. In region ${\rm UV}$ the two leading order contributions alone are integrated in Eq.~\eqref{eq:44}, since there are subleading contributions whose integrand is divergent as well, hindering convergence. In the following, letting the position of the cut ($k_{\rm UV}$) vary, we will show that such term can indeed be neglected.
\begin{figure}
  \includegraphics[width=0.5\linewidth]{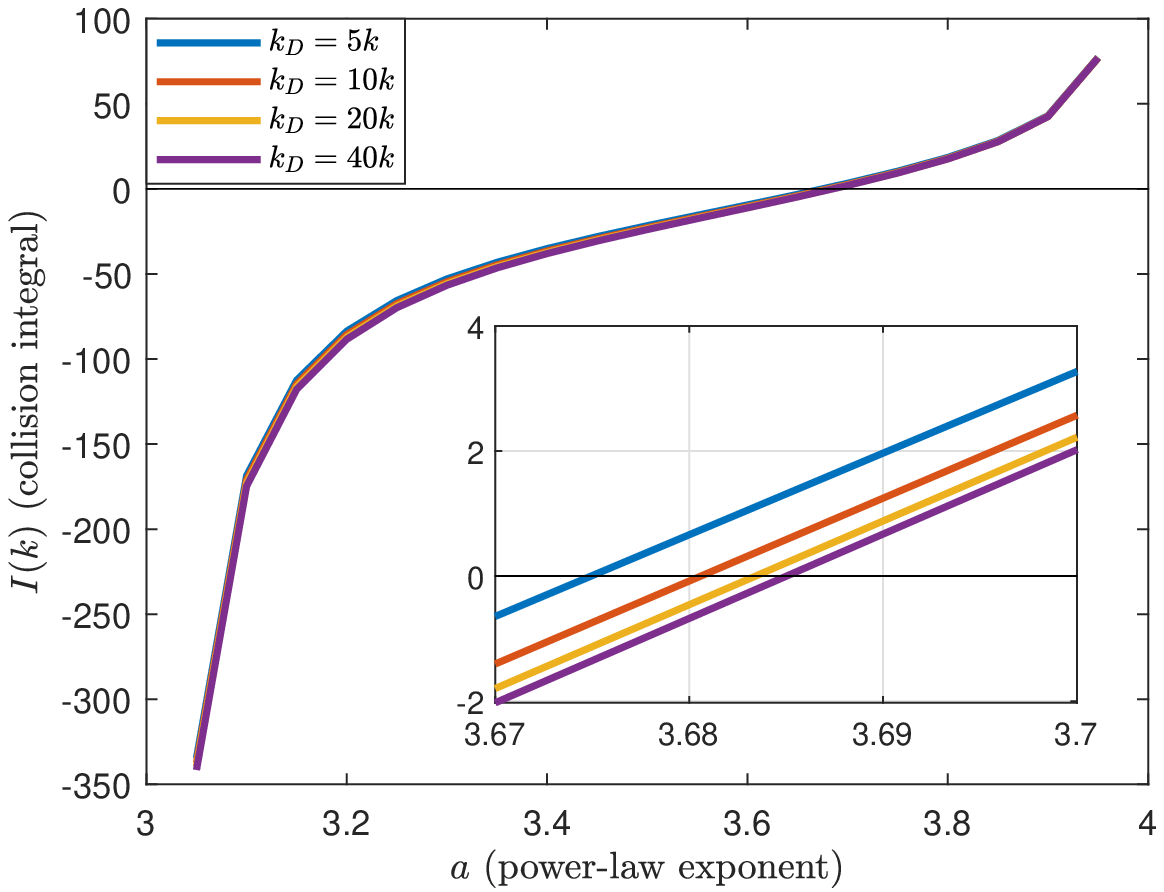}%
\includegraphics[width=0.5\linewidth]{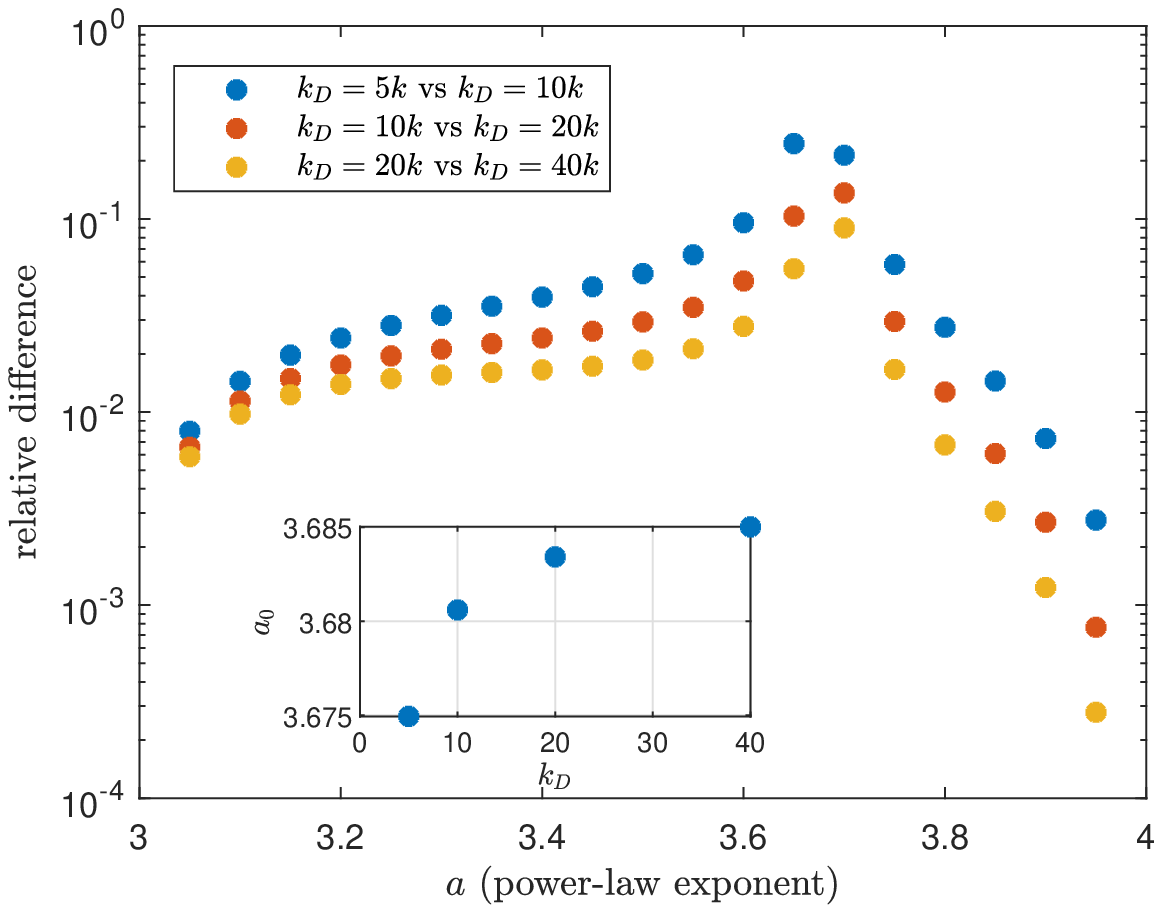}
  \caption{Left: Result of the integral as a function of the exponent $a$, as the cut at $k_1=k_{\rm UV}$ is sent toward infinity. The inset shows a zoomed region around the point where the integral vanishes. Right: Relative difference between the curves in the left panel, in logaritmic scale, giving evidence of convergence. In the inset, convergence of the stationary solution exponent (zero-crossing point in the left panel) is shown.
 \label{fig:5}}
\end{figure}
The width of the regions around $k_2=0$ is determined by the parameter $k_{\rm IR}$, while the cut at large $k$'s is performed at $k_1=k_{\rm UV}$. For the result to be general, it must be independent of the choice of $k_{\rm IR}$ and $k_{\rm UV}$, as long as they are finite numbers, $k_{\rm IR}$ being sufficiently small and $k_{\rm UV}$ sufficiently large. This has been checked. In Fig.~\ref{fig:5} we show how convergence is reached as $k_{\rm UV}$ increases, as the neglected contribution in ${\rm UV}$ vanishes. Independence of the result upon variations of $k_{\rm IR}$ is even more robust (not shown here).
\end{document}